\documentclass[10pt,conference]{IEEEtran}

\IEEEoverridecommandlockouts
\usepackage{hhline}
\usepackage{multirow}
\usepackage{color}
\usepackage{clrscode}
\usepackage{amsmath}
\usepackage{pifont}
\usepackage{subcaption} 
\usepackage[normalem]{ulem}
\usepackage{booktabs}
\usepackage{amsthm}
\usepackage{semantic}
\usepackage{url}
\usepackage[most]{tcolorbox}

\usepackage{graphicx}
\usepackage{xcolor,colortbl}
\usepackage{float}
\usepackage{array}
\usepackage{makecell}

\usepackage{hyperref}
\usepackage{enumitem}
\usepackage[most]{tcolorbox}
\usepackage{fancyvrb}
\usepackage{cleveref}

\PassOptionsToPackage{hyphens}{url}
\usepackage{cite}
\usepackage{xspace}
\def\BibTeX{{\rm B\kern-.05em{\sc i\kern-.025em b}\kern-.08em
    T\kern-.1667em\lower.7ex\hbox{E}\kern-.125emX}}

\newif\ifshowcomments
\showcommentstrue

\usepackage{array,collcell} 

\ifshowcomments
\newcommand{\wcp}[1]{\textcolor{brown}{[chengpeng: #1]}}

\else
\newcommand{\wcp}[1]{}
\fi

{\theoremstyle{definition}}
{\theoremstyle{definition}}
{\theoremstyle{definition}}
{\theoremstyle{definition}}
{\theoremstyle{definition}\newtheorem{example}{Example}}

\newcommand{\sysname}{{\sc RFCAudit}\xspace}
\usepackage{cite}
\usepackage{amsmath,amssymb,amsfonts}
\usepackage{algorithmic}
\usepackage{graphicx}
\usepackage{textcomp}
\usepackage{xcolor}
\usepackage[linesnumbered,ruled,vlined]{algorithm2e}
\def\BibTeX{{\rm B\kern-.05em{\sc i\kern-.025em b}\kern-.08em
    T\kern-.1667em\lower.7ex\hbox{E}\kern-.125emX}}
\begin{document}

\title{\sysname: AI Agent for Auditing Protocol Implementations Against RFC Specifications}

\author{
Mingwei Zheng\textsuperscript{1}, Chengpeng Wang\textsuperscript{1}, Xuwei Liu\textsuperscript{1}, Jinyao Guo\textsuperscript{1}, Shiwei Feng\textsuperscript{1},and Xiangyu Zhang\textsuperscript{1} \\
\textsuperscript{1}Department of Computer Science, Purdue University, West Lafayette, USA \\
Email: \{zheng618, wang6590, liu2598, guo846, feng292, xyzhang\}@purdue.edu}


\maketitle

\begin{abstract}
Functional correctness is critical for ensuring the reliability and security of network protocol implementations. Functional bugs, instances where implementations diverge from behaviors specified in RFC documents, can lead to severe consequences, including faulty routing, authentication bypasses, and service disruptions. Detecting these bugs requires deep semantic analysis across specification documents and source code, a task beyond the capabilities of traditional static analysis tools.
This paper introduces \sysname, an autonomous agent that leverages large language models (LLMs) to detect functional bugs by checking conformance between network protocol implementations and their RFC specifications. Inspired by the human auditing procedure, \sysname comprises two key components: an indexing agent and a detection agent. The former hierarchically summarizes protocol code semantics, generating semantic indexes that enable the detection agent to narrow down the scanning scope. The latter employs demand-driven retrieval to iteratively collect additional relevant data structures and functions, eventually identifying potential inconsistencies with the RFC specifications effectively.
We evaluate \sysname across six real-world network protocol implementations. \sysname identifies 47 functional bugs with 81.9\% precision, of which 20 bugs have been confirmed or fixed by developers.
\end{abstract}


\newcommand{\q}[1]{``{#1}''}

\newcommand{\code}[1]{\texttt{\small#1}\xspace}

\newcommand{\todoc}[2]{{\textcolor{#1}{\textbf{#2}}}}
\newcommand{\todoblack}[1]{{\todoc{black}{\textbf{[[#1]]}}}}
\newcommand{\todored}[1]{{\todoc{red}{\textbf{[[#1]]}}}}
\definecolor{applegreen}{rgb}{0.55, 0.71, 0.0} 
\newcommand{\todogreen}[1]{\todoc{applegreen}{\textbf{[[#1]]}}}
\newcommand{\todoblue}[1]{\todoc{blue}{\textbf{[[#1]]}}}
\newcommand{\todoorange}[1]{\todoc{orange}{\textbf{[[#1]]}}}
\newcommand{\todobrown}[1]{\todoc{brown}{\textbf{[[#1]]}}}
\newcommand{\todogray}[1]{\todoc{gray}{\textbf{[[#1]]}}}
\newcommand{\todopurple}[1]{\todoc{purple}{\textbf{[#1]}}}
\newcommand{\todopink}[1]{\todoc{magenta}{\textbf{[[#1]]}}}
\newcommand{\todocyan}[1]{\todoc{cyan}{\textbf{[[#1]]}}}
\newcommand{\todoviolet}[1]{\todoc{violet}{\textbf{[[#1]]}}}
\newcommand{\todoteal}[1]{\todoc{teal}{\textbf{[[#1]]}}}
\newcommand{\todo}[1]{\todored{TODO: #1}}

\definecolor{light-gray}{gray}{0.7}
\newcommand{\hilight}[1]{\colorbox{light-gray}{#1}}

\makeatletter
\newcommand*{\textoverline}[1]{$\overline{\hbox{#1}}\m@th$}
\makeatother

\newcommand\blackcircle[1]{%
  \tikz[baseline=(X.base)] 
    \node (X) [draw, shape=circle, inner sep=0, scale=0.8, fill=darkgray, text=white] {\strut #1};%
}

\newcommand\whitecircle[1]{%
  \tikz[baseline=(X.base)] 
    \node (X) [draw, shape=circle, inner sep=0, scale=0.8, fill=white, text=black] {\strut #1};%
}



\newif\ifenablecomments  
\enablecommentsfalse   

\ifenablecomments
    \newcommand{\mingwei}[1]{\todoblue{Mingwei: #1}}
    \newcommand{\danning}[1]{\todogreen{Danning: #1}}
    \newcommand{\wang}[1]{\todobrown{chengpeng: #1}}
    \newcommand{\xz}[1]{\todored{XZ: #1}}
\else
    \newcommand{\mingwei}[1]{}
    \newcommand{\danning}[1]{}
    \newcommand{\wang}[1]{}
    \newcommand{\xz}[1]{}
\fi

\newboolean{showchanges}
\setboolean{showchanges}{false} 

\newcommand{\change}[1]{%
    \ifthenelse{\boolean{showchanges}}%
        {\textcolor{blue}{#1}}
        {#1}
}


\newcounter{finding}
\newcommand{\intuition}[1]{
\begin{tcolorbox}[colback=gray!10, colframe=gray!20, boxrule=0pt, boxsep=2pt,
left=2pt, right=2pt, top=1pt, bottom=1pt, sharp corners, leftrule=2mm, leftrule=2pt, colframe=gray]
\refstepcounter{finding}
\textbf{Conclusion~\thefinding{}:} \emph{#1}
\end{tcolorbox}
}


\captionsetup[figure]{font=bf,skip=6pt}
\captionsetup[table]{font=bf,skip=6pt}
\newcommand{\distance}{8pt}
\setlength{\textfloatsep}{6pt}
\setlength{\floatsep}{\distance}
\setlength{\intextsep}{\distance}
\setlength{\dbltextfloatsep}{\distance} 
\setlength{\dblfloatsep}{\distance} 

\section{Introduction}
Network protocols are essential to digital communication, establishing standardized rules that govern data exchange between devices. They specify message formats and sequencing, as well as mechanisms for routing, connection management, and resource allocation.
Due to their complexity, implementations of network protocols are susceptible to functional bugs—logical errors that cause deviations from intended behaviors,
which undermines the reliability and security of network infrastructures.
For example, the Zerologon vulnerability (CVE-2020-1472~\cite{Zerologon}) in Microsoft's Netlogon Remote Protocol (MS-NRPC)~\cite{MS-NRPC} stemmed from improper initialization of a cryptographic vector, which allows attackers to impersonate any host, including domain controllers, and gain unauthorized domain administrator access. Due to its critical severity, it was assigned a maximum CVSS score of 10.0.

\begin{figure}[t]
\includegraphics[width=\linewidth]{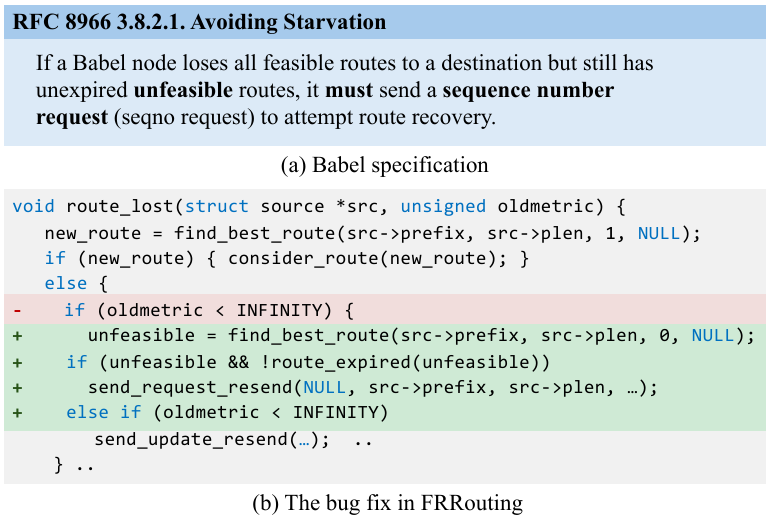}
\centering
\caption{An example functional bug in Babel and its fix}
\vspace{-1mm}
\label{fig:motiv}
\end{figure}

To enhance the reliability and security of network protocols, it is essential to identify functional bugs in their implementations.
Manually detecting such bugs is notoriously difficult, as developers have to compare the implementation against every relevant clause in lengthy RFC documents—a process that demands significant human effort and domain expertise.
However, automating functional bug detection remains a challenging task.
First, the semantic properties depicting protocol functionality are highly diverse and expressed informally in natural language within RFC documents.
In contrast to generic low-level safety properties~\cite{DBLP:journals/corr/AmorimHP17}, which can often be uniformly represented using logical constraints, application-specific high-level semantic properties~\cite{LTLFuzz, feng2024rocas, 
feng2025intentest, kate2025roscallbax} (or {\em semantic properties} in this paper), such as those found in network protocols, are considerably more difficult to formalize due to their reliance on domain knowledge and context-dependent behaviors.
As shown in the example in Figure~\ref{fig:motiv}(a),
such properties in the RFC document often require reasoning about the states of multiple protocol entities and the complex interactions among them,
making encoding such properties inherently intricate.
Second, the diversity of semantic properties significantly hinders the development of effective detection mechanisms.
Conventional bug detection techniques, such as constraint-based methods (e.g., symbolic execution~\cite{klee, 255310, 217563, SFA-Miner, TensileFuzz}) and graph-based reasoning approaches (e.g., data-flow reachability analysis~\cite{10.1145/3656400, 10.1145/3180155.3180178}), are inadequate for automatically validating the semantic properties of network protocols. These methods cannot automatically localize the functions relevant to specific properties and verify their compliance with the specification. For instance, it is challenging to localize the function \texttt{route\_lost} shown in Figure~\ref{fig:motiv}(b) and leverage a uniform reasoning framework to detect violations of the associated semantic property.

Although recent studies have proposed various techniques for bug detection in network protocol implementations, they remain inadequate to effectively identify functional bugs.
In general, existing approaches fall into three categories.
First, fuzzing techniques aim to uncover bugs by generating crash-triggering inputs~\cite{chatafl, mGPTFuzz, FuzzInMem}. However, functional bugs typically do not cause crashes or other observable anomalies, making them difficult for fuzzing-based methods to detect.
Second, differential analysis can reveal behavioral inconsistencies and discover potential functional bugs by comparing multiple independent implementations of the same protocol~\cite{Pardiff, ODIT, reen2020dpifuzz}. Yet, this approach depends on the availability of high-quality alternative implementations and does not guarantee conformance to the original specification, thereby limiting its recall for functional bug detection.
Third, formal verification offers strong semantic guarantees but requires rigorous formalization of protocol properties and formal reasoning frameworks~\cite{Pistachio, MusuvathiE04}. These requirements may not be easily achievable for real-world protocols due to the informal and diverse specifications highlighted above.
These limitations suggest the need for a new paradigm of bug detection that enables semantic reasoning on both informal specifications and implementation behavior, an essential capability for effectively detecting functional bugs.

This paper presents \sysname, the first LLM agent for detecting functional bugs in network protocol implementations.
Our approach is motivated by the observation that LLMs exhibit strong capabilities in understanding both natural language specifications, such as RFC documents, and the semantics of code snippets when they are well scoped, such as those spanning just a few related functions.
Notably, RFC documents are typically well-structured and articulate protocol specifications clearly in natural language. Meanwhile, protocol implementations often follow intuitive naming conventions that reflect the functionality of functions and modules.
This suggests a promising opportunity: Given an informal property described in an RFC, we can leverage LLMs to align the natural language description of the property with the code structure.
If the functions related to the property can be precisely identified by further code retrieval (e.g., along calling contexts), LLMs can perform comprehensive and accurate reasoning over these functions to discover potential functional inconsistencies, thereby enabling functional bug detection.
Technically, \sysname consists of two collaborative agents, namely an \emph{\textbf{indexing agent}} and a \textbf{\emph{detection agent}}, which perform the following core analyses, respectively.

\begin{itemize}[leftmargin=0.3cm]
\item \textit{Code Semantic Indexing.}
To bridge informal properties in RFC documents with relevant code scope, the indexing agent constructs hierarchical semantic indexes of the protocol implementation using LLMs,
which summarizes the semantics of functions, files, and directories into concise natural language descriptions.
Such semantic indexes allow \sysname to narrow down the scope of functions relevant to the target property, significantly reducing token and time costs during the detection. 
\vspace{1mm}

\item \textit{Retrieval-Guided Detection.}
To collect sufficient contexts,
the detection agent progressively retrieves additional relevant functions using symbolic tools driven by the LLM.
If the currently available context suffices to validate or refute conformance to the semantic property, \sysname proceeds to the next property. Otherwise, it augments the context by retrieving callers or callees until a conclusion can be drawn.
The retrieved functions can also support self-critics in the detection, during which the detection agent reviews its own reasoning steps upon the retrieved functions to mitigate hallucinations and improve the precision of the detection.
\end{itemize}

We implement \sysname as a prototype~\cite{rfctool}, powered by Claude 3.5 Sonnet. We apply \sysname to the implementations of six widely-used network protocols, including Babel, DHCP, and IGMP, and assess its ability to detect functional bugs according to the corresponding RFC documents.
It is shown that \sysname successfully identifies 47 zero-day functional bugs with a high precision of 81.9\%. 
In comparison, existing advanced agents, such as Copilot powered with three advanced models: Claude 3.7 Sonnet, Claude 3.5 Sonnet, and GPT-4o, only identify 26 bugs on average, while exhibiting a substantially higher average false positive rate of 77.5\%.
We further conduct comprehensive ablation studies, which highlight the significant precision gains attributed to the designs of code semantic indexing and retrieval-guided detection. In summary, our work makes the following contributions:
\begin{itemize}[leftmargin=0.3cm]
    \item We propose \sysname, the first LLM agent that checks the compliance of network protocol implementation and RFC documents for functional bug detection.

    \item We introduce a multi-agent design that summarizes code functionalities as semantic indexes and leverages them for demand-driven code retrieval, facilitating functional bug detection with high precision.

    \item We evaluate \sysname on six network protocol implementations and identify 47 functional bugs with 81.9\% precision, outperforming existing LLM-based baselines in both precision and semantic coverage. 20 of the detected bugs have been confirmed or fixed by the respective developers. 
\end{itemize}

\section{Overview}
\label{sec:motiv}
In this section, we begin with a real-world functional bug in a network protocol implementation
and discuss the limitations of existing approaches (\Cref{subsec:example}).
Then we provide the key ideas of our approach (\Cref{subsec:approach}) with illustrative examples.

\subsection{Motivating Example}\label{subsec:example}

\begin{figure*}[t]
\includegraphics[width=0.9\textwidth]{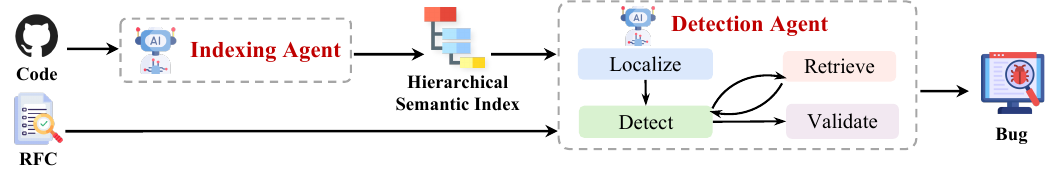}
\centering
\vspace{-2mm}
\caption{The workflow of \sysname}
\vspace{-1mm}
\label{fig:workflow}
\end{figure*}

\Cref{fig:motiv} shows a real-world functional bug in FRRouting, a collaborative project under the Linux Foundation with over 400 contributors, powering major open-source large-scale networking platforms like Microsoft’s SONiC and Amazon’s DENT. According to the RFC document shown in \Cref{fig:motiv}(a), when a node loses all feasible routes to a destination but still holds unexpired unfeasible routes, it MUST send a sequence number (seqno) request to trigger route recovery.
An unfeasible route means a neighbor recently advertised a path that is not currently usable, but might become valid again. Sending a seqno request prompts that neighbor to re-announce the route, potentially restoring connectivity.
However, the original implementation fails to check for unfeasible routes, which can leave the router in a starved state with no way to recover, causing persistent connectivity loss in dynamic networks. 
As shown in \Cref{fig:motiv}(b), the fix adds an explicit check for unexpired unfeasible routes and issues a seqno request when such a route is found, ensuring RFC compliance and preventing persistent routing failure. This bug has been confirmed and fixed by the developers.

However, detecting functional bugs is extremely challenging~\cite{Pardiff, KIT}.
Traditional bug detection methods, such as fuzzing~\cite{pham2020aflnet, chatafl, shi2023lifting} and static analysis~\cite{shi2018pinpoint, Xue16SVF}, primarily focus on identifying low-level issues like memory corruption. Consequently, these approaches are ineffective against non-crashing protocol non-compliance, such as the bug illustrated in \Cref{fig:motiv}. 
Model checking techniques also encounter significant difficulties in detecting this bug, primarily due to their dependence on explicitly specified and formally defined properties. For instance, tools such as CMC~\cite{MusuvathiE04} and UPPAAL~\cite{DiazCRP04} efficiently verify high-level behaviors like state transitions and message sequencing, while struggling to capture diverse and nuanced semantic properties, exemplified in \Cref{fig:motiv}(a). Moreover, even when semantic properties can be explicitly specified, model checking requires extensive manual abstraction efforts, which are often labor-intensive and error-prone, constraining their practical applicability. 
Lastly, recent studies such as ParDiff~\cite{Pardiff} and ParCleanse~\cite{ParCleanse} target non-crashing parsing bugs by inferring valid packet formats from code or RFCs.
Although effective in identifying parser-related issues, these approaches fall short in detecting functional bugs that involve more complex semantic properties in components beyond network packet parsers.

\vspace{-1mm}
\subsection{\sysname in a Nutshell}\label{subsec:approach}

To bridge the gap, we propose \sysname, an LLM-powered autonomous agent that identifies semantic inconsistencies between source code and RFC documents. Our approach is motivated by the observation that LLMs, having been pre-trained on vast corpora of natural language and programming code, exhibit strong capabilities in understanding both the specification offered by the RFC document and program semantics in the code. For instance, given the semantic property shown in Figure~\ref{fig:motiv}(a) and the buggy implementation in Figure~\ref{fig:motiv}(b), LLMs can potentially identify the inconsistency between the RFC documents and the corresponding implementation logic.
However, applying LLMs for functional bug detection in network protocols is far from trivial, requiring us to resolve two critical technical challenges.

\begin{figure*}[t]
\includegraphics[width=0.9\textwidth]{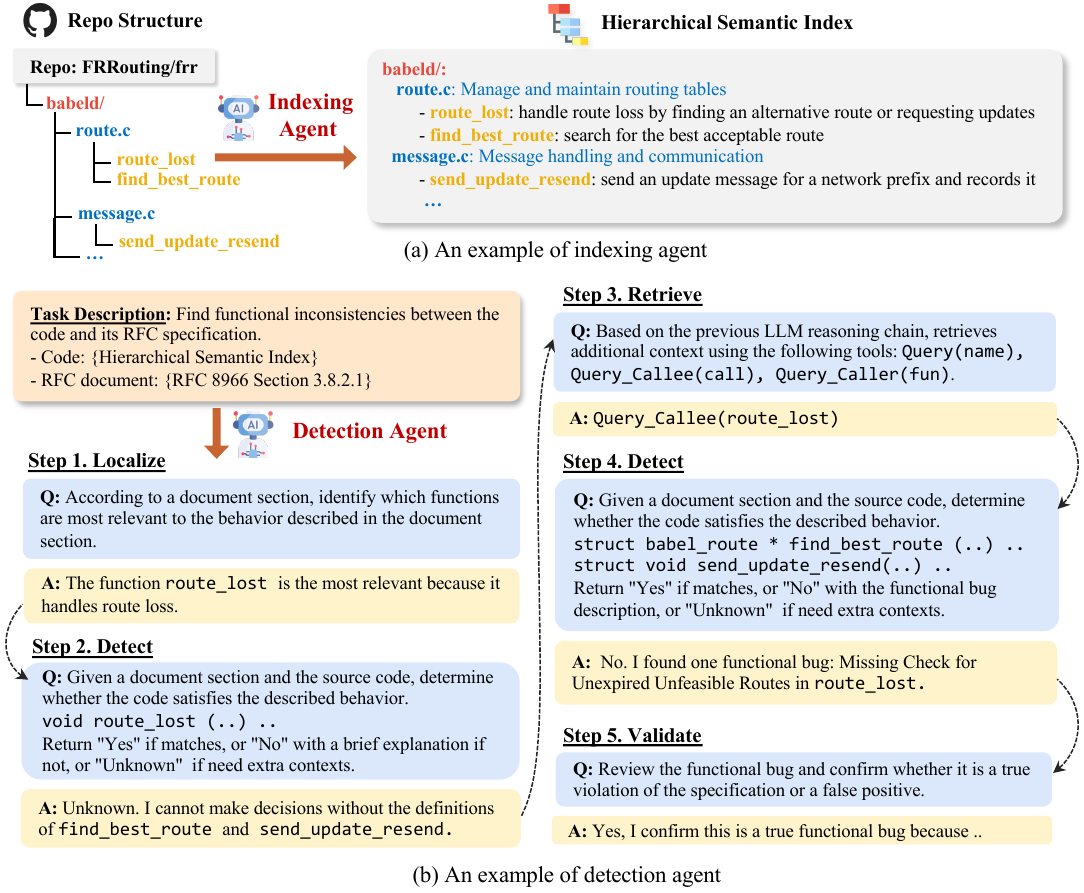}
\centering
\vspace{-2mm}
\caption{Illustrative examples of the two agents in \sysname}
\label{fig:workflow_example}
\end{figure*}

\begin{itemize}[leftmargin=0.3cm]
  \item 
  Due to the lack of direct correspondence between RFC segments and source code, identifying where a high-level property is realized in the implementation is quite difficult. For instance, checking the property in Figure~\ref{fig:motiv}(a) requires finding the related function \texttt{route\_lost} in Figure~\ref{fig:motiv}(b), which involves exploring a large codebase without clear guidance from the RFC document.

  \item 
  Even after locating a relevant function, LLMs may still struggle to determine whether the implementation satisfies the intended property. Since a semantic requirement is often implemented through multiple interconnected functions spread across files or modules, LLMs typically need additional contexts, such as the functions \texttt{find\_best\_route} and \texttt{send\_update\_resend} in \Cref{fig:motiv}.
\end{itemize}

To tackle these challenges, we draw inspiration from the way human developers typically audit functional correctness against protocol specifications. When verifying whether a particular semantic property is correctly implemented, experienced developers first perform a preliminary walkthrough of the codebase to become familiar with the roles of different functions, files, and modules. 
 They then progressively gather additional program constructs, such as related functions or data structures, to reason about the correctness of the implementation and identify potential violations.
 
\sysname mirrors this human auditing process through two key technical components: \emph{\textbf{code semantic indexing}} and \emph{\textbf{retrieval-guided detection}}. 
The former creates a semantic map of the codebase by summarizing the functionality of functions, files, and directories in natural language, allowing fast and informed localization of potentially relevant code. The latter incrementally augments the reasoning context with related program constructs, enabling the system to validate a property when its implementation spans multiple functions or modules.
These two techniques are realized through two collaborative agents in \sysname, the \emph{\textbf{indexing agent}} and the \emph{\textbf{detection agent}}, whose respective workflows are illustrated in \Cref{fig:workflow}.

\begin{itemize}[leftmargin=0.3cm]
  \item \emph{\textbf{Indexing Agent.}} The indexing agent performs semantic summarization and generates \emph{hierarchical semantic indexes} at the directory, file, and function levels. As shown in Figure~\ref{fig:workflow_example}(a), for the Babel protocol, files like \texttt{route.c} and \texttt{message.c}, as well as functions like \texttt{route\_lost}, \texttt{find\_best\_route}, and \texttt{send\_update\_resend}, are annotated with concise summaries of their functionality. These semantic descriptions are later utilized by the detection agent to navigate and localize relevant code.

  \item \emph{\textbf{Detection Agent.}} 
  Given a semantic property, the detection agent uses hierarchical indexes to identify inconsistencies. As shown in \Cref{fig:workflow_example}(b), it proceeds with four actions in five steps: (1) Use the semantic indexes to localize a relevant function, such as \texttt{route\_lost}; (2) Detect inconsistencies based on this context; (3) Retrieve additional relevant functions, such as \texttt{find\_best\_route} and \texttt{send\_update\_resend}, if the context is insufficient; (4) Draw a conclusion about a potential violation; and (5) Apply a self-critique strategy to reassess the reasoning for hallucination mitigation.
\end{itemize}

\begin{figure*}
\includegraphics[width=\textwidth]{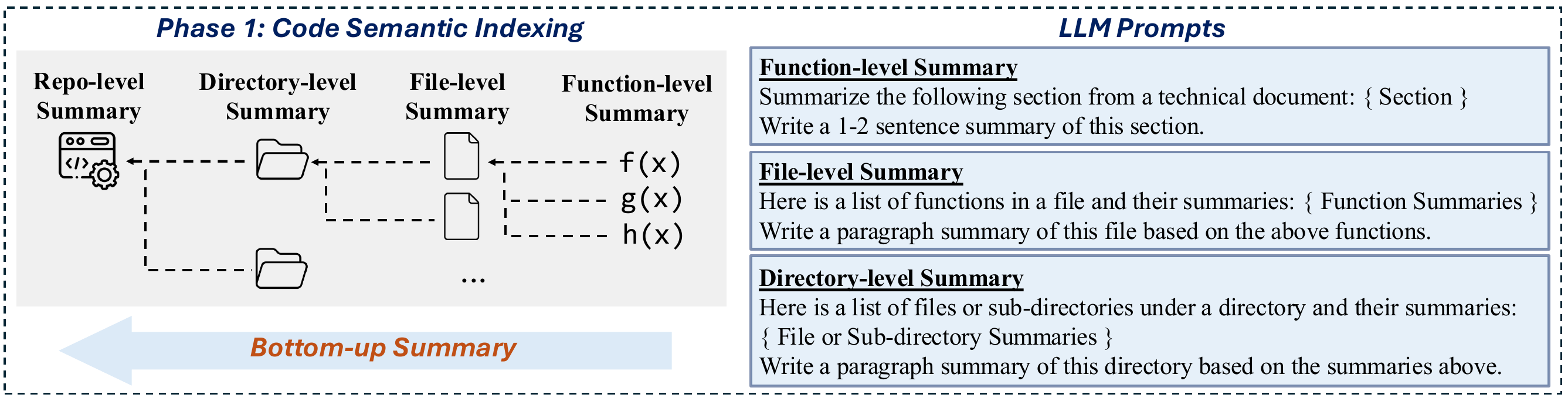}
\centering
\caption{The workflow of code semantic indexing and prompt templates}
\label{fig:index}
\end{figure*}

\section{Our Approach}
\label{sec:approach}

\sysname emulates the reasoning process of human developers when auditing for functional bugs and consists of two collaborative agents, the indexing agent and the detection agent, which support scalable and effective functional bug detection upon network protocol implementations.
In what follows, we present more technical details of the two agents with concrete prompts and illustrative examples.

\subsection{Phase 1: Code Semantic Indexing}
\label{subsec:stage1}

To bridge RFC documents with relevant code segments, we follow the behavior of human developers by constructing semantic indexes over the protocol implementation. This approach enables us to capture the semantics of the code in a manner analogous to the way developers form a high-level understanding of the implementation.
Basically, a simple indexing design is to utilize the names of functions, files, and directories as summaries, as they often suggest the intended functionality~\cite{chen2025locagent,xia2024agentless}. 
However, such names are not always informative enough to capture precise semantics. For example, in the Babel implementation shown in Figure~\ref{fig:motiv}, both \texttt{babeld.c} and \texttt{babel\_main.c} reside in the same directory, yet their names do not clearly reveal their distinct functionalities. To generate more informative semantic indexes, we leverage LLMs to summarize the contents of functions, files, and directories, producing concise natural language descriptions that form a hierarchical semantic view of the implementation.

\vspace{-1mm}
As shown in Figure~\ref{fig:index}, the indexing agent constructs hierarchical semantic indexes in a bottom-up manner. It begins at the \emph{function level}, where each function definition is passed to the LLM with a prompt, which asks for a concise summary of its semantics. At the \emph{file level}, summaries of all functions within the same source file are aggregated to prompt the LLM to generate a description of the file's overall functionality, capturing how individual behaviors contribute to the role of the file in the protocol implementation.
This process continues at the \emph{directory level}, where the agent combines summaries of all files and subdirectories to generate a higher-level description of the directory. Finally, summaries from all directories are aggregated into a repository-level summary, which is a top-level directory summary, reflecting the system’s overall structure and responsibilities.
This bottom-up indexing process produces a structured semantic index aligned with the code hierarchy, enabling precise and efficient retrieval during detection. The corresponding prompt templates for each indexing level are shown on the right of Figure~\ref{fig:index}.
Notably, the code semantic indexing is one-time effort. When the codebase changes, we can incrementally update the indexes. In our evaluation, we choose the Claude 3.5 Sonnet model and the average cost for a full repository is \$1.89, shown in \Cref{tab:cost}. We can also consider using cheaper models for indexing as the task is relatively simple and does not rely on strong reasoning ability.

\begin{example}
Consider the repository structure shown on the left of \Cref{fig:workflow_example}(a).
We first ask the LLMs to summarize each function.
For example, \texttt{route\_lost} handles route loss,
\texttt{find\_best\_route} searches for the best acceptable route,
and \texttt{send\_update\_resend} sends and records update messages.
Next, the function-level summaries under the same file are combined to produce a file-level summary that captures the main responsibility of each file. In this case,  we can discover that \texttt{route.c} focuses on route maintenance, and \texttt{message.c} on inter-node communication. 
Finally, all file summaries are merged into a directory-level summary for the \texttt{babeld} module.
Lastly, we can obtain the hierarchical semantic indexes shown on the right of \Cref{fig:workflow_example}(b).
\end{example}

\begin{figure}
\includegraphics[width=\linewidth]{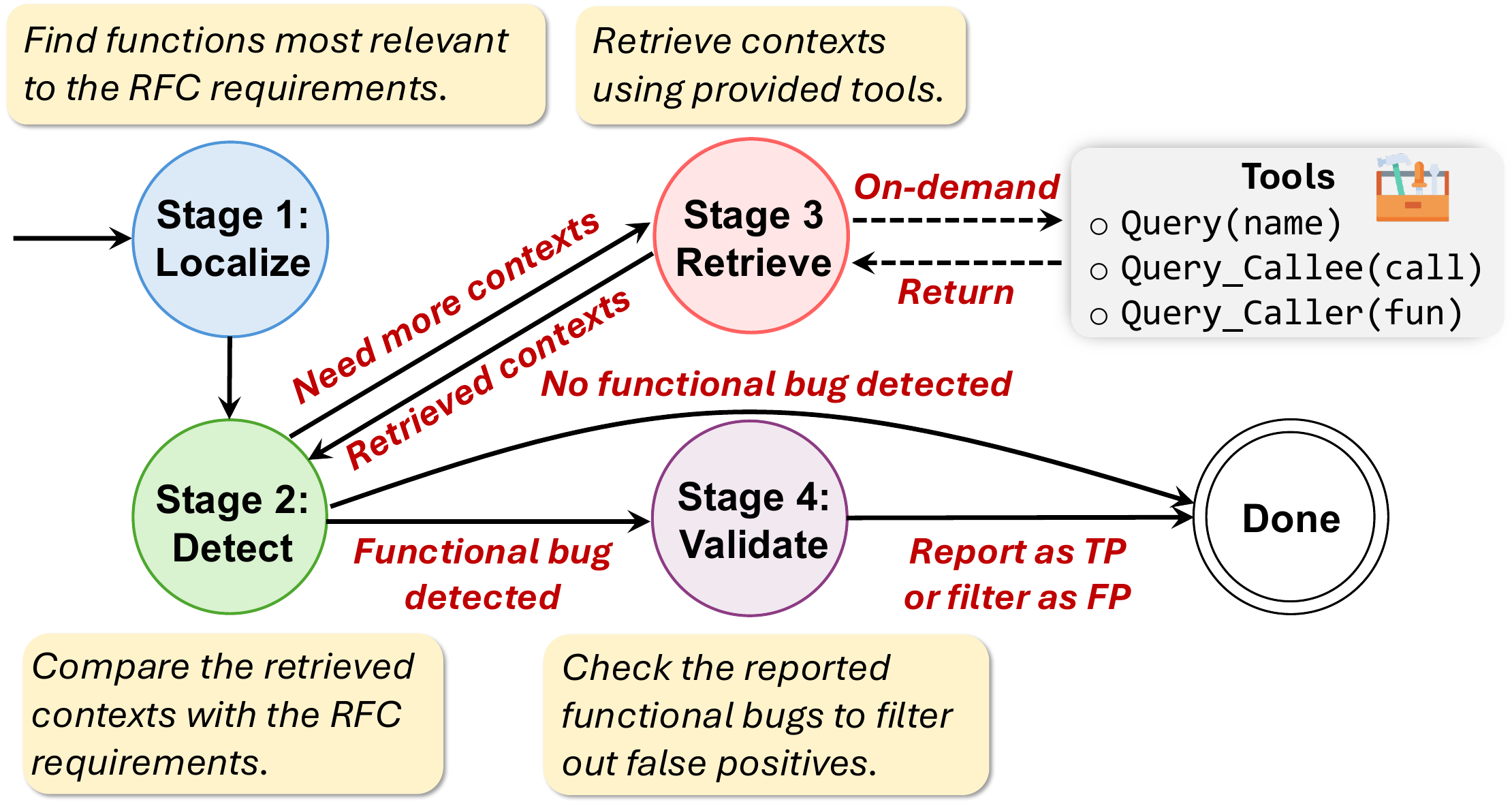}
\centering
\caption{The state machine of the detection agent
}
\label{fig:fsm}
\end{figure}

\begin{figure*}[t!]
\includegraphics[width=\linewidth]{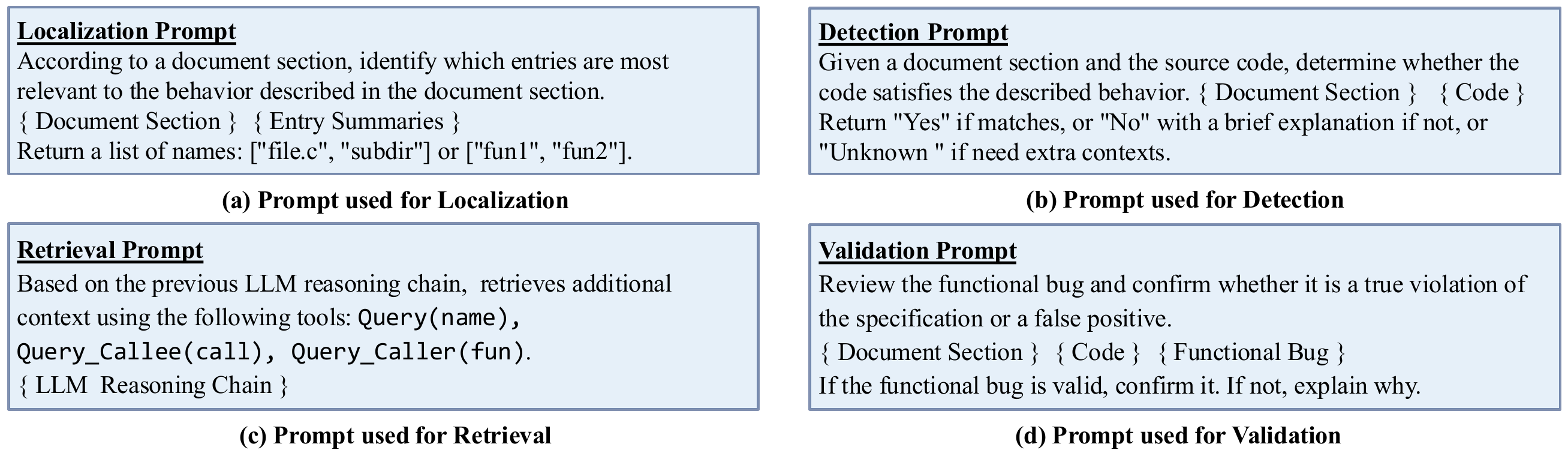}
\centering
\caption{The prompt templates used in the detection agent
} 
\vspace{-1mm}
\label{fig:prompts}
\end{figure*}

\subsection{Phase 2: Retrieval-guided Detection}
\label{subsec:stage2}

As demonstrated in Section~\ref{subsec:stage1}, the hierarchical semantic indexes allow us to effectively bridge RFC documents with their corresponding implementation functions. Similar to how human developers audit code, the analysis often requires collecting additional program constructs, such as caller/callee functions and data structure definitions, to augment the context until a reliable conclusion can be drawn regarding the conformance or violation of a given semantic property.
Building on this insight, we introduce the detection agent, which performs retrieval-guided detection to identify inconsistencies between the RFC specification and the code as functional bugs. 

Technically, the retrieval-guided detection begins by pre-processing the RFC document and segmenting it into sections and subsections based on its structural headings. From each subsection, the LLMs extract mandatory semantic properties, which serve as guides for the detection process. A single subsection may yield multiple properties depending on its content.
For each semantic property, the detection agent scans the implementation by following a workflow similar to a finite state machine, as illustrated in Figure~\ref{fig:fsm}. Specifically, it first localizes relevant functions for inspection (\textbf{\textit{Localization}}), determines whether the implementation satisfies the RFC requirement (\textbf{\textit{Detection}}), and, if necessary, retrieves additional context (\textbf{\textit{Retrieval}}). When a potential functional bug is found, the detection agent examines the reasoning chain that produces the reported bug via self-criticism (\textbf{\textit{Validation}}).
In the following subsections, we provide further details on each component of the detection workflow shown in Figure~\ref{fig:fsm}.

\smallskip
\noindent
\emph{\textbf{{Localization.}}}
Guided by the pre-built hierarchical semantic indexes, the detection agent first identifies functions that are likely responsible for implementing the behavior described in a given RFC section. This localization is performed in a top-down manner. Starting from the root directory of the protocol implementation, it navigates through different levels of directories and files by instantiating the prompt template shown in Figure~\ref{fig:prompts}(a).
At each level, it reviews the semantic summaries of directories or files, selects those most relevant to the RFC section, and descends into the selected entries. Upon reaching a file, it examines the summaries of its functions and identifies those that align with the described behavior. Because this search process is recursive, the final set of relevant functions may span multiple files across the codebase.

\smallskip
\noindent
\emph{\textbf{{Detection.}}}
Based on the localized relevant functions, the detection agent attempts to identify inconsistencies between the implementation and the RFC. Using the template in Figure~\ref{fig:prompts}(b), it queries the LLM to make one of three decisions, as shown in Figure~\ref{fig:fsm}.
Specifically, if the LLM detects a potential functional bug, the agent proceeds to validate it (see the \emph{Validation} stage below). If the LLM concludes that the implementation conforms to the RFC requirement, the agent terminates the analysis for the current property. Otherwise, the current context is insufficient, the agent initiates an additional retrieval to expand the context (see the \emph{Retrieval} stage below).
Notably, the detection agent resembles the reasoning pattern of human developers, who iteratively gather evidence from the program constructs to identify potential bugs or to justify the correctness of an implementation.

\smallskip
\noindent
\emph{\textbf{{Retrieval.}}}
To assist the LLM in identifying functional bugs or justifying correctness, the detection agent enables demand-driven retrieval of additional program constructs. This is achieved by equipping the LLM with a set of predefined tools, which it can invoke through function calls as needed.
Specifically, the agent provides three types of tools. First, \texttt{Query(name)} retrieves the definition of a data structure and a macro. This is useful when the current context references symbols whose definitions are not yet available.
Second, \texttt{Query\_Callee(call)} returns the definition of the callee function invoked at the specified call site.
This is used when the current function doesn’t provide enough information to determine whether the behavior matches the RFC, and the LLM suspects that the called function contains important logic.
This selective strategy enables the LLM to focus on semantically meaningful functions while avoiding unnecessary expansion of trivial or unrelated calls.
Third, \texttt{Query\_Caller(fun)} retrieves all callers of the function \texttt{fun}, allowing the LLM to examine how the function is used, particularly whether its preconditions are satisfied by the callers.
These retrieval operations are triggered on demand, guided by the LLM’s reasoning. Benefiting from model's planning ability, the LLMs can choose the proper tools for retrieval. The retrieved results further augment the analysis context in the \textit{Detection} stage for continued analysis. The LLM prompt is shown in Figure~\ref{fig:prompts}(c).

\begin{figure}[t]
\includegraphics[width=0.9\linewidth]{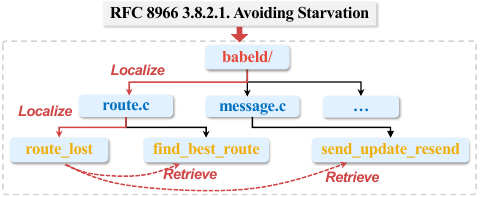}
\centering
\caption{An illustrative example of Phase 2} 
\vspace{-1mm}
\label{fig:detection_example}
\end{figure}

\smallskip
\noindent
\emph{\textbf{{Validation.}}}
Due to the potential lengthy detection context, the LLMs may hallucinate and report a false positive.
To mitigate hallucination, we employ self-critics~\cite{lin-etal-2024-criticbench} in the detection agent.
Specifically, the detection agent reviews all available information, including the RFC section, the retrieved context, and earlier reasoning steps, to make a final judgment.
If the violation is confirmed, the functional bug is reported as a true positive. Otherwise, it is discarded as a false positive.
It may also uncover bugs missed during previous detection.
The prompt is instantiated using the template in \Cref{fig:prompts}(d). Our results in \Cref{sub:ablation} show that the self-critics can reduce the false positives by 71.7\% (from 63.9\% to 18.1\%).

\begin{example}
Given the RFC section in \Cref{fig:motiv}(a) and the semantic indexes in \Cref{fig:workflow_example}(a), the detection agent identifies the relevant function \texttt{route\_lost} in \texttt{route.c} (\Cref{fig:detection_example}). Finding its context insufficient, the agent retrieves \texttt{find\_best\_route} and \texttt{send\_update\_resend}, which select candidate routes and issue reactive updates, respectively. Notably, \texttt{send\_update\_resend} resides in a different file. With this broader context, the agent observes that only feasible routes are searched (\texttt{flag = 1}) and no sequence number request is made in the fallback logic, violating the starvation prevention rule. Retrieval stops here, and the bug is finally identified as shown in \Cref{fig:motiv}. The whole process is detailed in \Cref{fig:workflow_example}(b).
\end{example}

\section{Implementation and Evaluation}\label{sec:eval}
\sysname is implemented atop AutoGen~\cite{AutoGen}, a multi-agent framework designed for building LLM applications. To extract functional specifications, we collect RFC documents in plain text format and derive informal functional properties through structured segmentation of the textual content. We employ Claude 3.5 Sonnet~\cite{claude} as the underlying LLM for \sysname, configured with a temperature of 0.0 to enforce greedy decoding. This setting reduces randomness in model responses, thereby enhancing the consistency and reproducibility of \sysname. To facilitate the indexing and detection agents, we implement analysis tools using Tree-sitter~\cite{treesitter}, a parsing library for different programming languages. 
In particular, we construct call graphs by leveraging function names and the numbers of their associated parameters or arguments. 
The output generated by the indexing agent is stored in a JSON file, enabling incremental analysis as the protocol implementation evolves. This design avoids redundant processing by skipping the analysis of unchanged functions, files, and modules.

To assess the performance of \sysname, we conduct experiments to address the following research questions:
\begin{itemize}[leftmargin=0.1cm]
    \item \textbf{RQ1}: How effective does \sysname identify functional bugs in real-world network protocol implementations?
    \item \textbf{RQ2}: How does \sysname\ compare to existing approaches?
    \item \textbf{RQ3}: What are the runtime and token costs of \sysname?
    \item \textbf{RQ4}: How do the two agents contribute to performance?  
\end{itemize}

\vspace{-2mm}
\begin{table}[t]
\centering
\caption{The statistics of evaluation subjects.}
\label{tab:evaluation-subjects}
\resizebox{\linewidth}{!}{%
\begin{tabular}{l|l|c|c}
 \toprule
\multirow{1}*{\textbf{Name}} & \textbf{Description}  & \textbf{LoC} & \textbf{Version (\#Page)}\\
\midrule
Babel & Distance-vector Routing Protocol &9.6K & RFC 8966 (54) \\
BFD &Bidirectional Forwarding Detection Protocol & 17.3K  & RFC 5880 (49)\\
NHRP & NBMA Next Hop Resolution Protocol & 9.2K & RFC 2332 (52)\\
RIPng  & Routing Information Protocol Next Generation &9.3K  & RFC 2080 (19)\\
DHCP & Dynamic Host Configuration Protocol & 8.6K  & RFC 2131 (45)\\
IGMP & Internet Group Management Protocol & 5.2K  & RFC 2236 (24)\\
\bottomrule
\end{tabular}
}
\end{table}

\subsection{Dataset}
We evaluate \sysname on six protocol implementations from two widely-used, open-source network stacks: FRRouting~\cite{frr} and lwIP~\cite{lwip}.
FRRouting (FRR) is an Internet routing protocol suite for Linux and Unix platforms, with over 3.6K stars on GitHub. It is deployed in production by major infrastructure providers such as NVIDIA and Orange, and is distributed through mainstream Linux repositories including Debian and Fedora.
lwIP is a lightweight TCP/IP stack designed for resource-constrained environments, with over 1.3K stars on GitHub. It is widely integrated into embedded systems, IoT platforms, and real-time applications, and has been adopted by vendors like Intel, Xilinx, and Freescale.
Overall, both FRRouting and lwIP are actively maintained and widely used in practice, making them ideal subjects for assessing \sysname across diverse deployment contexts.
For each protocol, we collect the corresponding RFC that the implementation is based on as the reference. 
The protocol implementation sizes range from 5.2 KLoC to 17.3 KLoC, whereas the corresponding RFC documents span between 19 and 54 pages. The detailed information is shown in \Cref{tab:evaluation-subjects}.

\begin{table}[t]
\centering
\caption{The statistics of main results.}
\label{tab:inconsistencies}
\resizebox{\linewidth}{!}{%
\begin{tabular}{l|c|c|c|c|c|c}
\toprule
\multirow{2}{*}{\textbf{Protocols}} &\multirow{2}{*}{\textbf{\#Incons.}}& \multirow{2}{*}{\textbf{TP}} &\multirow{2}{*}{\textbf{Precision}}  & \multicolumn{2}{c|}{\textbf{Unique Bug}} & \multirow{2}{*}{\textbf{\#Total Bug}}\\
\cmidrule(lr){5-6}
&& & &\textbf{\#New} & \textbf{\#Old} \\
\midrule
Babel & 16 & 12 & 75.0\% & 8 & 2 & 10\\
BFD   & 30 & 28 & 93.3\%  & 9 & 5 & 14\\
NHRP  & 12 & 10 & 83.3\% & 10& 0 & 10\\
RIPng & 3  & 2  & 66.7\% &2 & 0 & 2\\
DHCP  & 12 & 8  & 66.7\% & 4& 0 &4\\
IGMP  & 10 & 8  & 80.0\% & 7& 0 &7\\
\midrule
Total & 83 & 68 & 81.9\% & 40 & 7 & 47\\
\bottomrule
\end{tabular}
}
\end{table}

\subsection{RQ1: Effectiveness of \sysname}

\subsubsection{The Effectiveness of Bug Detection}
\label{subsec:setup}
To answer RQ1, we apply \sysname to each protocol implementation and its corresponding RFC to detect inconsistencies. For every reported issue, we manually verify whether it represents a real violation, and record the counts of true and false positives to compute precision.
Since a single underlying issue may result in multiple reported bugs (e.g., when the same property is described across multiple sections of the RFC), we group relevant reports into unique bugs. For each unique bug, we check the project's issue tracker and pull requests to determine whether it is a previously unreported bug (i.e., new) or a known issue (i.e., old) that has not yet been fixed in the latest version.

\begin{figure}[t]
    \centering
    \includegraphics[width=0.83\linewidth]{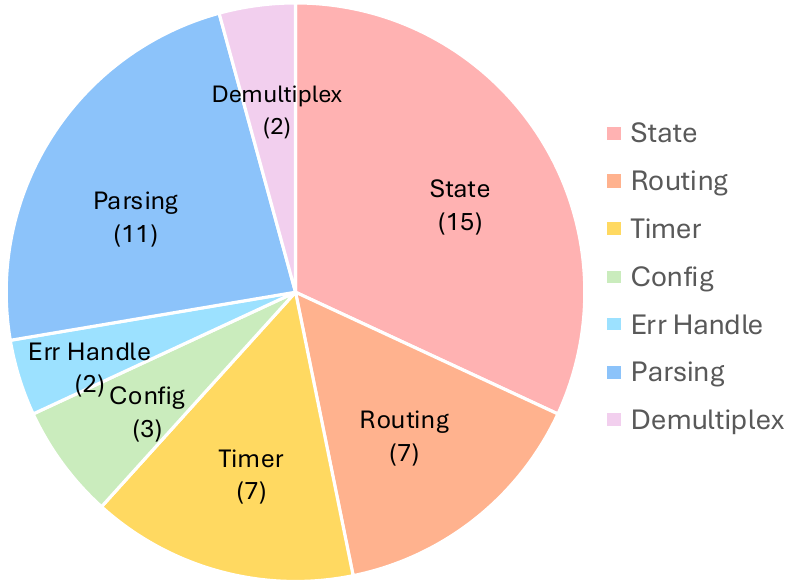}
    \caption{Distribution of functional bug categories.}
    \label{fig:bug-pie}
\end{figure}

\begin{table*}[t]
\centering
\small
\caption{The precision, recall and F1 score of context retrieval. NA indicates that the divisor is 0.}
\label{tab:context-completeness}
\resizebox{\linewidth}{!}{%
    \begin{tabular}{l|c|c c c|c c c|c}
      \toprule
      \multirow{2}{*}{\textbf{Protocol}} & \multirow{2}{*}{\textbf{Properties}}
        & \multicolumn{3}{c|}{\textbf{After \textit{Localization}}} 
        & \multicolumn{3}{c|}{\textbf{After \textit{Retrieval}}} 
        & \multirow{2}{*}{\textbf{\#Miss}}\\
      \cmidrule(lr){3-5} \cmidrule(lr){6-8}
        & & \textbf{Function (\%)} & \textbf{Type (\%)} & \textbf{Macro (\%)}
            & \textbf{Function (\%)} & \textbf{Type (\%)} & \textbf{Macro (\%)} & \\
      \midrule
      Babel & 15 & 50.0 / 91.7 / 64.7 & NA / 0.0 / NA & NA / 0.0 / NA & 50.0 / 91.7 / 64.7 & 100.0 / 100.0 / 100.0 & 100.0 / 100.0 / 100.0 & 1\\
      BFD  & 22 & 57.1 / 66.7 / 61.5 & NA / 0.0 / NA & NA / 0.0 / NA & 72.7 / 100.0 / 84.2 & 66.7 / 100.0 / 80.0 & 100.0 / 100.0 / 100.0 & 0\\
      NHRP & 15 & 71.4 / 85.7 / 77.9 & NA / 0.0 / NA & NA / 0.0 / NA & 76.9 / 95.2 / 85.1 & 100.0 / 100.0 / 100.0 & 100.0 / 100.0 / 100.0 & 0\\
      RIPng & 24 & 60.7 / 70.8 / 65.4  & NA / 0.0 / NA & NA / 0.0 / NA & 37.3 / 100.0 / 54.3 & 100.0 / 33.3 / 50.0 & NA / 0.0 / NA & 0\\
      DHCP & 34 & 60.7 / 68.0 / 64.1 & NA / NA / NA & NA / NA / NA & 65.3 / 94.0 / 77.1 & NA / NA / NA & NA / NA / NA & 0\\
      IGMP & 21 & 52.9 / 75.0 / 62.0 & NA / NA / NA & NA / 0.0 /  & / 100.0 / & NA / NA / NA & 100.0 / 100.0 / 100.0 & 0\\
      \midrule
      \textbf{Total} & 131 & 59.3 / 71.3 / 65.6 & NA / 0.0 / NA  & NA / 0.0 / NA & 56.0 / 95.1 / 70.5 & 87.5 / 77.8 / 82.4 & 100.0 / 75.0 / 100.0 & 1\\
      \bottomrule
    \end{tabular}%
  }
\end{table*}

\textbf{Results.} 
As shown in \Cref{tab:inconsistencies}, \sysname generates a total of 83 bug reports across six protocols, of which 68 are true positives, yielding an overall precision of 81.9\%.
Grouping related bug reports, we uncovered 47 unique bugs, including 40 new bugs and 7 already reported ones.
To date, 20 out of 47 unique bugs have been confirmed or their patches have been approved or merged by the developers.
Overall, these results demonstrate that \sysname is effective in identifying bugs across diverse protocols. 
To better understand the diversity of functional bugs detected by \sysname, we categorize the 47 unique bugs into seven categories, namely \textit{Parsing}, \textit{Config}, \textit{Routing}, \textit{Err Handle}, \textit{State}, \textit{Timer}, and \textit{Demultiplex}. \change{As shown in \Cref{fig:bug-pie}}, most bugs fall under State (15/47) and Parsing (11/47), reflecting common pitfalls in protocol state management and input validation.
\sysname also uncovered Timer misconfiguration (e.g., incorrect IGMP delays), Demultiplexing issues (e.g., incorrect BFD session lookup), and Err Handle gaps (e.g., miss error-on-error prevention). This variety highlights the ability of \sysname to detect diverse semantic violations specified by the RFCs. Full bug details are available in our artifact~\cite{rfctool}.

\subsubsection{The Effectiveness of Context Retrieval}
\change{
To assess context retrieval accuracy, we manually construct ground-truth mappings for each RFC property. Each property is annotated with the minimal set of functions, types, and macros required to capture its semantics. Since this annotation is labor-intensive, we sample 30 RFC subsections (5 per protocol) from our datasets, covering a total of 131 properties across six protocols. For each property, we compare the elements retrieved by \sysname against the ground truth and measure precision, recall, and F1 score after two stages: (i) \textit{localization} and (ii) \textit{retrieval}. We then compute average recall across all sampled properties to assess how effectively \sysname collects the necessary context for inconsistency detection.
Since both the retrieved element count and ground truth element count can be zero for types and macros, making recall undefined, we use ``NA\text{''} to indicate these cases. 
Finally, we examine missed bugs and identify those caused by incomplete retrieval.}

\textbf{Results.}  
Table~\ref{tab:context-completeness} reports the context retrieval accuracy across protocols. 
In the \textit{localization} stage, \sysname retrieves 172 functions, of which 102 are true positives, yielding 59.3\% precision and 71.3\% recall (102 of 143 ground-truth functions). This stage is limited to function retrieval and, by design, does not include types or macros. After the \textit{retrieval} stage, precision is 56.0\% for functions (136 of 243), 87.5\% for types (7 of 8), and 100.0\% for macros (7 of 7).
More importantly, recall improves significantly—95.1\% (136 of 143) for functions, 77.8\% (7 of 9) for types, and 75.0\% (6 of 8) for macros—with only one bug missed overall, caused by code misinterpretation rather than missing context.
In this task, recall is more critical than precision. Low precision only means that unnecessary elements are retrieved, causing additional overhead, while low recall means the needed context is completely missed,  preventing the detection of true inconsistencies. Note that achieving high precision is challenging for two reasons: (1) some functions are semantically close to the target property but not truly relevant, making them difficult to filter out, and (2) a single RFC property is often distributed across multiple functions or files, so aggressive filtering risks discarding essential information. For these reasons, our retrieval design explicitly prioritizes recall, even at the cost of including borderline or loosely related items. Precision-oriented refinements are left for future work.

\subsubsection{False Positive Analysis}
\label{subsec:fp}
We analyze the 15 false positives reported by \sysname to understand their root causes. Based on this analysis, we classify them into four categories: \emph{(i) incomplete context retrieval} (3 cases), where necessary function, type or macro definitions are not retrieved; \emph{(ii) RFC misinterpretation} (6 cases), where \sysname mistakenly treats optional or incorrect properties as mandatory requirements; \emph{(iii) code misinterpretation} (5 cases), where \sysname retrieves the correct context but fails to correctly interpret the code’s control flow or semantics; and \emph{(iv) intentional implementation choice} (1 case), where the observed behavior is a deliberate deviation from the RFC. The most common root cause is RFC misinterpretation, including 4 cases where the LLM extracts incorrect properties and 2 cases where the LLM treats optional clauses as mandatory. Such errors may be mitigated by fine-tuning the LLM on RFC texts to improve its ability to extract correct mandatory requirements.

\subsection{RQ2: Baseline Comparison}
\change{We compare \sysname with GitHub Copilot and LTL-Fuzzer~\cite{LTLFuzz}.
We do not compare with other network protocol bug detection tools, as they either cannot detect functional bugs (e.g., ChatAFL~\cite{chatafl}, NetLifter~\cite{shi2023lifting}), or target narrow scope like parsing (e.g., ParCleanse~\cite{ParCleanse}, ParVAL~\cite{ParVal}).}
\subsubsection{Comparison with GitHub Copilot}
We compare \sysname with GitHub Copilot using three LLMs: Claude 3.7 Sonnet (Thinking Mode), Claude 3.5 Sonnet, and GPT-4o. 
All are accessed via Copilot's Workspace Chat interface in VS Code, which supports repository-level queries.
To match \sysname{}’s design, we provide Copilot with one RFC section per prompt and ask it to identify deviations from implementations.
For Claude 3.7 Sonnet, we manually examine all generated bug reports.
However, Claude 3.5 Sonnet and GPT-4o produce hundreds of candidates, making full manual validation impractical.
To avoid over-reporting and minimize false positives, we instruct each LLM baseline to only report bugs with 100\% confidence.
To enable a fair and scalable evaluation, we adopt a sampling-based evaluation. For each protocol, we randomly sample up to twice the number of bug reports identified by \sysname (e.g., since \sysname generates 16 bug reports for Babel, we sample 32). If the model produces fewer than that, we evaluate all of them. Each sampled case is manually reviewed by a domain expert.

\begin{table*}[t]
  \centering
  \small
\caption{Baseline comparison across three model configurations.}
  \label{tab:baseline-merged}
  \resizebox{\textwidth}{!}{%
    \begin{tabular}{l|ccccc|ccccc|ccccc}
      \toprule
      \multirow{2}{*}{\textbf{Protocols}}
        & \multicolumn{5}{c|}{\textbf{Copilot + Claude 3.7 Sonnet}}
        & \multicolumn{5}{c|}{\textbf{Copilot + Claude 3.5 Sonnet (With Sample)}}
        & \multicolumn{5}{c}{\textbf{Copilot + GPT-4o (With Sample)}} \\
      \cmidrule(lr){2-6} \cmidrule(lr){7-11} \cmidrule(lr){12-16}
        & \textbf{\#Incons.} & \textbf{\#TP} & \textbf{Precision} & \textbf{\#New} & \textbf{\#Old}
        & \textbf{\#Incons.} & \textbf{\#Sample} & \textbf{Precision} & \textbf{\#New} & \textbf{\#Old}
        & \textbf{\#Incons.} & \textbf{\#Sample} & \textbf{Precision} & \textbf{\#New} & \textbf{\#Old} \\
      \midrule
      Babel  &  76 & 16 & 21.1\% &  7 &  3
             & 173 & 32 & 12.5\% & 3 & 1
             & 493 & 32 &  3.1\% & 1 & 0 \\
      BFD    &  45 & 37 & 82.2\% &  5 &  5
             & 115 & 60 & 23.3\% & 7 & 4
             & 244 & 60 & 20.0\% & 6 & 3 \\
      NHRP   &  27 & 10 & 37.0\% &  9 &  0
             &  89 & 24 & 20.8\% & 4 & 0
             & 204 & 24 & 16.7\% & 4 & 0 \\
      RIPng  &   6 & 1 & 16.7\% &  1 &  0
             &  45 & 6 & 16.7\% & 1 & 0
             &  84 & 6 &  0.0\% & 0 & 0 \\
      DHCP   &  32 & 9 & 28.1\% &  4 &  0
             &  64 & 24 & 25.0\% & 4 & 0
             & 188 & 24 & 42.0\% & 1 & 0 \\
      IGMP   &  29 & 5 & 17.2\% &  3 &  0
             &  52 & 20 & 15.0\% & 2 & 0
             & 102 & 20 &  5.0\% & 1 & 0 \\
      \midrule
      \textbf{Total}
             & 215 & 78 & \textbf{36.3\%} & 29 &  8
             & 538 & 166 & \textbf{19.9\%} & 21 & 5
             & 1315 & 166 & \textbf{11.4\%} & 13 & 3 \\
      \bottomrule
    \end{tabular}%
  }
\end{table*}

\begin{table*}[t]
\centering
\caption{Token usage (In: input tokens, Out: output tokens), financial cost, and execution time per protocol.}
\label{tab:cost}
\resizebox{\textwidth}{!}{%
\begin{tabular}{l|ccc|ccc|ccc}
\toprule
\multirow{2}{*}{\textbf{Protocol}} & \multicolumn{3}{c|}{\textbf{Phase 1: Code Semantic Indexing}} & \multicolumn{3}{c|}{\textbf{Phase 2: Retrieval-guided Detection}} & \multicolumn{3}{c}{\textbf{Total}} \\
\cline{2-10}
\addlinespace[0.3em]
& \textbf{Tokens (In/Out)} & \textbf{Cost (\$)} & \textbf{Time (min)} 
& \textbf{Tokens (In/Out)} & \textbf{Cost (\$)} & \textbf{Time (min)} 
& \textbf{Tokens (In/Out)} & \textbf{Cost (\$)} & \textbf{Time (min)} \\
\midrule
Babel & 261K / 71K & 1.85& 44& 1340K / 41K & 4.63 & 39 & 1061K / 112K & 6.48&83\\
BFD   & 449K / 119K & 3.13 &71 & 1086K / 30K & 3.71 & 52 & 1535K / 149K & 6.84 & 123\\
NHRP  & 319K / 81K & 2.17 & 49 & 960K / 25K  & 3.26 & 56 & 1279K / 106K & 5.43 & 105\\
RIPng & 197K / 52K & 1.37 & 31 & 429K / 9K  & 1.42 & 23 & 626K / 61K & 2.79 & 54\\
DHCP  & 248K /44K& 1.40 & 28 & 1353K / 24K & 4.42 & 47 & 1601K / 68K & 5.82& 75\\
IGMP  & 248K / 28K & 1.40 & 28& 390K / 16K  & 1.41 & 20 & 638K / 44K &2.81 &48\\
\midrule
\textbf{Average} & 287K / 66K & 1.89 & 42 & 926K / 24K & 3.14 & 39 & 1123K / 90K& 5.03 & 81\\
\bottomrule
\end{tabular}
}
\end{table*}

\textbf{Results.} 
Table \ref{tab:baseline-merged} shows the performance of GitHub Copilot under three model configurations.
With Claude 3.7 Sonnet (no sampling), it generated 215 bug reports, of which 78 were true positives (36.3\% precision), uncovering 37 unique bugs (29 new, 8 old).
Claude 3.5 Sonnet achieved 19.9\% precision over 538 inconsistencies and found 26 bugs.
GPT-4o returned the most bug reports (1,315) but the lowest precision (11.4\%) and the fewest bugs (16).
Overall, all three baselines suffer from low precision and high validation overhead. In contrast, \sysname achieves 81.9\% precision and discovers 47 bugs in total (40 new) with far less manual effort (Table \ref{tab:inconsistencies}). \sysname consistently reports the lowest false-positive rates and the highest bug counts across all protocols. Notably, \sysname’s use of Claude 3.5 (a non-reasoning model) still outperforms Claude 3.7 (a reasoning model).

\subsubsection{Comparison with LTL-Fuzzer}
\change{We compare \sysname with LTL-Fuzzer~\cite{LTLFuzz}, which checks protocol implementations against LTL properties via instrumentation and fuzzing.
Setting up LTL-Fuzzer requires substantial manual efforts: translating informal specifications into LTL formulas and mapping each atomic predicate to code locations. Consequently, it mainly captures event-ordering temporal properties but cannot naturally express data-format checks, arithmetic constraints, or timing requirements. We applied \sysname to the 15 zero-day bugs reported in LTL-Fuzzer’s evaluation and also examined whether the 47 violated properties from our dataset could be encoded for LTL-Fuzzer properties.}

\textbf{Results.} \sysname successfully detected 12 of the 15 zero-days originally reported by LTL-Fuzzer, without requiring manual property construction or predicate-to-location mapping. These included 7 in TinyDTLS and 5 in Contiki-Telnet. In contrast, when we attempted to encode the 47 violated properties from our dataset as LTL-Fuzzer properties, only 11 could be expressed; the remaining 36 required constraints beyond plain LTL or lacked precise predicate locations. These results indicate that \sysname is effective in automatically detecting temporal property violations while extending to a broader range of functional inconsistencies.

\subsection{RQ3: Efficiency of \sysname}
\textbf{Setup and Metrics.}
To quantify the efficiency of \sysname, we measure input/output token usage, financial cost, and execution time for analyzing each repository. The results are reported separately for Phase 1 (Code Semantic Indexing) and Phase 2 (Retrieval-guided Detection). Although Claude 3.5 is used for both phases in our evaluation, Phase 1 primarily involves lightweight summarization. Since it does not require complex reasoning, it could be replaced by smaller or less expensive LLMs with minimal impact on performance.

\textbf{Results.} As shown in \Cref{tab:cost}, \sysname uses an average of 1123K input and 90K output tokens, costs \$5.03, and completes each protocol analysis in 81 minutes. Since the analysis has a complete summarization of the whole code base and compares code against the full RFC, input token usage and cost largely depend on the code base size and RFC length. Overall, \sysname is efficient for functional bug detection.

\subsection{RQ4: Ablation Studies}\label{sub:ablation}
\textbf{Ablation Study 1: Without Code Semantic Indexing.}
To evaluate the impact of code semantic indexing, we disable semantic summarization in Phase 1. The agent still follows the directory structure and explores the codebase hierarchically but no longer uses LLM-generated summaries. Instead, it relies solely on directory and file names plus function signatures. This setup follows the same design of AGENTLESS~\cite{xia2024agentless} and LocAgent~\cite{chen2025locagent}, which uses code structures without semantic summaries.
As shown in \Cref{tab:ablation}, removing semantic indexing drops precision from 81.9\% to 51.4\% and reduces detected bugs from 47 to 26.
To explain this drop, we analyze each false positive and classify its root cause according to the four categories in \Cref{subsec:fp}. We observe that false positives due to \emph{(i) incomplete context retrieval} rise significantly (from 3 to 14), making it the main source of false positives in this setting. Without semantic summaries, the agent confuses similarly named helper functions across modules—for example, retrieving \texttt{parse\_update\_subtlv} instead of the correct \texttt{parse\_packet} when validating Babel update TLVs.
This highlights the importance of semantic summarization for accurate context retrieval and reduced false positives.

\smallskip

\textbf{Ablation Study 2: Detection Without Retrieval.}
To evaluate the impact of the \textit{Retrieval} stage, we disable all code retrieval tools (\texttt{Query}, \texttt{Query\_Callee}, \texttt{Query\_Caller}). The detection agent must make decisions using only the initially retrieved function(s), without fetching additional context. As shown in Table~\ref{tab:ablation}, this results in a sharp drop in precision (from 81.9\% to 47.1\%) and fewer detected bugs (from 47 to 36). This shows that retrieval is important for reducing false positives and improving detection quality.
To assess the impact of each retrieval tool, we disable each tool individually. As shown in \Cref{tab:ablation2}, removing any single tool sharply reduces precision from 81.9\% with full retrieval to about 47–50\%, with little difference across tools. However, the number of detected bugs varies, with the largest drop occurring when \texttt{Query\_Caller} is removed (from 47 to 39), suggesting it contributes the most to overall bug coverage.

\begin{table}[t]
  \centering
  \small
  \caption{Ablation Studies.}
  \label{tab:ablation}
  \resizebox{\linewidth}{!}{%
    \begin{tabular}{l|c@{\hskip 5pt}c@{\hskip 5pt}c|c@{\hskip 5pt}c@{\hskip 5pt}c|c@{\hskip 5pt}c@{\hskip 5pt}c}
      \toprule
      \multirow{2}{*}{\textbf{Protocol}} 
        & \multicolumn{3}{c|}{\textbf{Ablation Study 1}} 
        & \multicolumn{3}{c}{\textbf{Ablation Study 2}} 
        & \multicolumn{3}{c}{\textbf{Ablation Study 3}} \\
      \cmidrule(lr){2-4} \cmidrule(lr){5-7} \cmidrule(lr){8-10}
        & \textbf{Precision} & \textbf{\#New} & \textbf{\#Old} 
        & \textbf{Precision} & \textbf{\#New} & \textbf{\#Old}
        & \textbf{Precision} & \textbf{\#New} & \textbf{\#Old}\\
      \midrule
      Babel  & 52.6\% & 5 & 3 & 48.0\% & 9 & 3 &  35.2\% & 11 & 3 \\
      BFD    & 68.8\% & 8 & 5 & 70.6\% & 5 & 5 &  67.9\% & 10 & 5\\
      NHRP   & 20.0\% & 2 & 0 & 45.0\% & 6 & 2 & 20.4\% & 7 &0\\
      RIPng  & NA & 0 & 0 & 25.0\% & 1 & 0 & 5.9\% & 1 & 0\\
      DHCP   & 0\% & 0 & 0& 50.0\% & 3 & 0  & 30.0\% & 4& 0\\
      IGMP   & 57.1\% & 3 & 0  & 18.2\% & 2 & 0 & 29.6\% & 4& 0\\
      \midrule
      \textbf{Total} &\textbf{51.4\%} & 18 & 8 & \textbf{47.1\%} & 26 & 10  & \textbf{36.1\%} & 37 & 8 \\
      \bottomrule
    \end{tabular}%
  }
\end{table}

\begin{table}[t]
  \centering
  \small
  \caption{Impact of Removing Single Retrieval Tool.}
  \label{tab:ablation2}
  \resizebox{\linewidth}{!}{%
    \begin{tabular}{ccc|ccc|ccc}
      \toprule
      \multicolumn{3}{c|}{\textbf{w/o \texttt{Query}}} 
        & \multicolumn{3}{c}{\textbf{w/o \texttt{Query\_Callee}}} 
        & \multicolumn{3}{c}{\textbf{w/o \texttt{Query\_Caller}}} \\
      \cmidrule(lr){1-3} \cmidrule(lr){4-6} \cmidrule(lr){7-9}
        \textbf{Precision} & \textbf{\#New} & \textbf{\#Old} &
        \textbf{Precision} & \textbf{\#New} & \textbf{\#Old} &
        \textbf{Precision} & \textbf{\#New} & \textbf{\#Old}\\
      \midrule
      \textbf{50.0\%} & 31 & 10 & \textbf{47.7\%} & 37 & 8 & \textbf{47.3\%} & 31 & 8 \\
      \bottomrule
    \end{tabular}%
 } 
\end{table}
\smallskip
\textbf{Ablation Study 3: Detection Without Self-critics.}
To evaluate the impact of self-critics, we disable it during detection. This leads to a higher false positive rate (18.1\% to 63.9\%) and fewer bugs (47 to 45), emphasizing its crucial role in enhancing detection precision and finding overlooked bugs.

\begin{figure}[t]
\includegraphics[width=\linewidth]{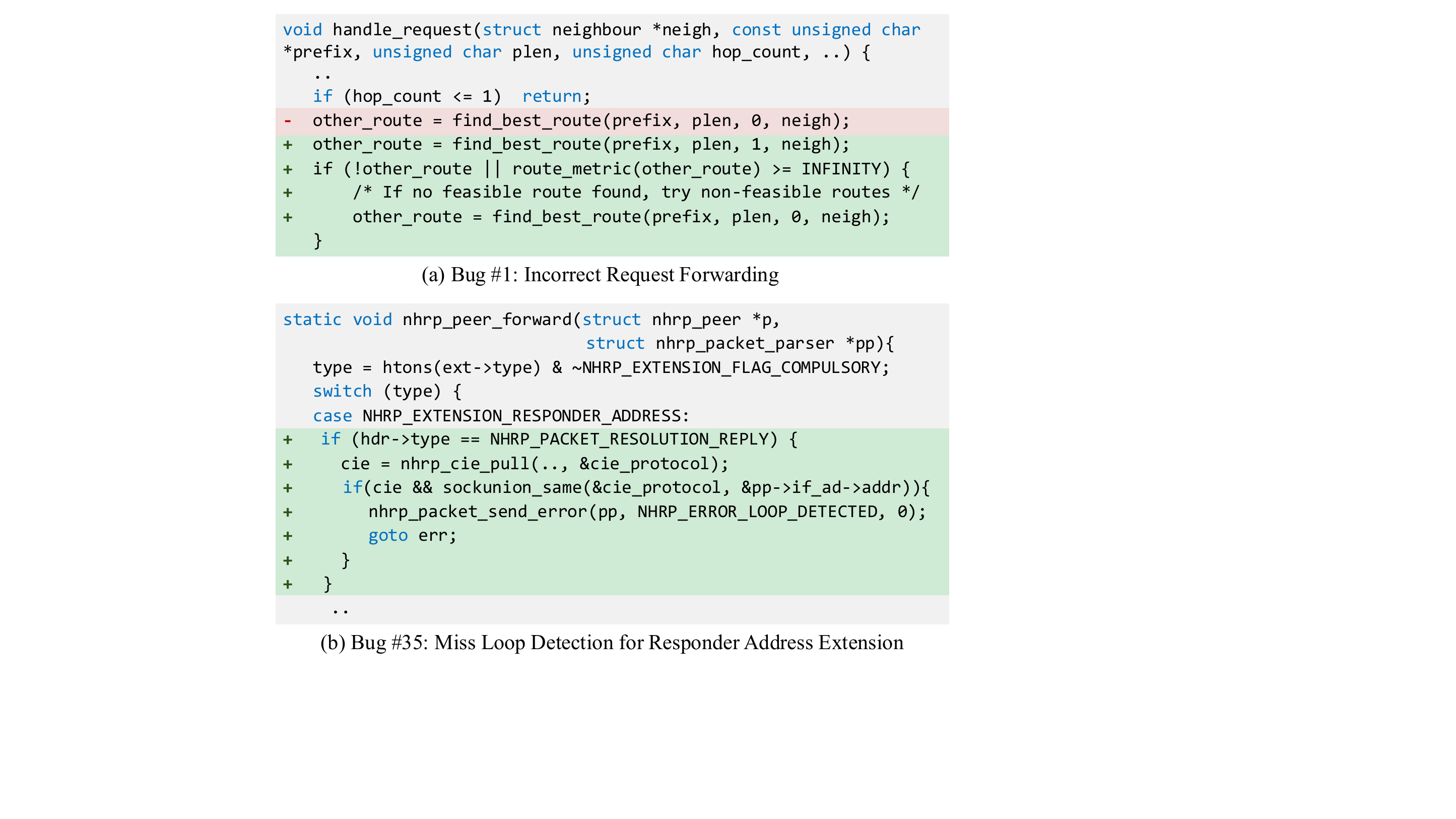}
\centering
\vspace{-5mm}
\caption{Case Studies.}
\label{fig:case}
\vspace{-2mm}
\end{figure}

\subsection{Case Studies}\label{subsec:casestudy}
We present two representative case studies:
\smallskip

\textbf{Bug \#1: Incorrect request forwarding}.
RFC 8966 specifies an ordered forwarding rule: when hop count is 2 or more, a node must first attempt to forward via a feasible route (excluding the requester); only if no such route exists should it fall back to a non-feasible one. 
As shown in \Cref{fig:case}(a), FRRouting ignored this ordering and always chose a non-feasible route. This misbehavior could cause forwarding loops or prevent requests from reaching valid next hops, undermining routing convergence. We submitted a fix to prioritize feasible routes, which has been merged by the developers.

\smallskip

\textbf{Bug \#35: Miss loop detection}.
RFC 2332 specifies that if an NHS forwards an NHRP Resolution Reply that lists its own protocol address in the Responder Address Extension, it must discard the packet and generate an NHRP Error Indication of type “Loop Detected”. As shown in \Cref{fig:case}(b), FRRouting omitted this check, allowing persistent forwarding loops that consume resources and delay address resolution. The bug was confirmed by the developers.

 \subsection{Threats to Validity.}\label{subsec:threats}

Several threats may affect the validity of our findings. 
First, we adopt Claude 3.5 Sonnet as our language model with a temperature setting of 0.0 for stable analysis results, thereby enhancing the reproducibility of \sysname. Nevertheless, alternative models or configurations may yield variations in performance~\cite{deng2025swe, xie2023impact, xie2025core}. 
So we ran controlled experiments using two additional general-purpose API models: GPT-4o and DeepSeek-V3, to analyze the BFD protocol in FRRouting, where \sysname discovered the most bugs.
\sysname detected 11 and 13 true bugs with the precision of 71.4\% and 70.3\%, respectively,
demonstrating the effectiveness across models.
Second, our manual classification of true positives and false positives may introduce subjective bias. To mitigate this risk, two authors independently reviewed each reported functional bug and resolved any disagreements through discussion~\cite{xie2023impact, 9678600}. We further validated our results by reporting newly discovered bugs to protocol developers, where all bug reports that received a response have been confirmed as true positives.
Third, our method relies on high-quality and well-structured documentation such as RFCs. Ambiguous or poorly maintained documentation might hinder semantic understanding, potentially threatening the effectiveness of our technique. 

\subsection{Evaluation on Industry-Scale Protocol Stacks}\label{subsec:larger}

We further applied \sysname to two large protocol stacks: \textbf{aws-c-http}, a 74K LoC HTTP library in the AWS Common Runtime, and the \textbf{Linux TCP/IP stack}, about 190K LoC in the kernel implementing core transport and network protocols.
We evaluated aws-c-http against RFC 7230 (HTTP/1.1), and Linux against RFC 793 (TCP), RFC 768 (UDP), and RFC 791 (IP). On aws-c-http, \sysname detected 18 inconsistencies, of which 12 were confirmed as true positives, yielding 12 unique bugs. On the Linux TCP/IP stack, it flagged 3 inconsistencies, with 1 true positive. These results demonstrate that \sysname scales to industry-scale codebases and can expose real specification violations with high precision.
\section{Discussion and Future Work}\label{sec:discuss}
\change{While our current evaluation targets network protocols, RFCAudit is not limited to this domain. The key requirement is the availability of specification documents that define expected functionality and constraints. In domains such as system libraries, cloud APIs, or security-critical frameworks, specifications are often available as API references, design guidelines, or standards documents. \sysname can be adapted to align such specifications with code and flag inconsistencies as potential bugs. We choose network protocols for evaluation because RFCs are detailed and publicly accessible, but extending RFCAudit to other domains is a promising next step to further demonstrate its generality and effectiveness.}

\section{Related Work}\label{sec:related}

\subsection{Bug Detection in Network Protocol Implementation}
The correctness of network protocol implementations is critical for ensuring secure and reliable digital communication. Existing bug detection techniques for protocol implementations can be broadly categorized into three approaches.
First, fuzzing-based techniques—such as BooFuzz~\cite{boofuzz} and SAGE~\cite{Msrsage}—primarily identify bugs by triggering crashes during execution. Netlifter~\cite{shi2023lifting} and ChatAFL~\cite{chatafl} improve fuzzing coverage by using static analysis or LLMs to infer packet formats and generate valid inputs, but still rely on crashes, overlooking subtle semantic issues. LLMIF~\cite{wang2024llmif} and mGPTFuzz~\cite{mGPTFuzz} extract formats or FSMs from specifications using LLMs to uncover semantic bugs, but focus narrowly on state transition or input validation violations and are limited to specific IoT devices, hard to generalize to broader protocol testing.
\change{LTL-Fuzzer~\cite{LTLFuzz} uses LTL formulas to guide fuzzing and detect event-ordering violations, but demands significant manual effort for formula crafting and predicate-to-code mapping, and is limited to temporal property violations.}
Second, differential analysis techniques~\cite{Pardiff, ODIT, reen2020dpifuzz} detect bugs by comparing the behaviors of multiple independent implementations of the same protocol. These approaches can uncover semantic discrepancies that fuzzing may miss. However, they are inherently limited to scenarios where multiple protocol implementations are available for comparison.
Third, formal verification-based techniques rigorously check protocol implementations against formally specified properties. While powerful in reasoning about semantic correctness, they require substantial manual effort to construct, maintain, and validate formal models, posing significant barriers to practical deployment.
Beyond the three categories, EBugDec~\cite{EBugDec} targets inconsistencies between RFC evolution and protocol implementations. However, its scope is limited to RFC-evolutionary bugs, primarily focusing on packet parsing issues.
In contrast to these prior efforts, our proposed technique, \sysname, addresses functional correctness by bridging the informal specifications in RFC documents with the program semantics of the protocol implementation, facilitating the functional bug detection with high precision and efficiency.

\subsection{LLM-aided Static Bug Detection}
LLM-aided static bug detection has advanced rapidly~\cite{DBLP:journals/pacmse/WangZW024, ye2024threads, DBLP:journals/corr/abs-2405-17238, yang2025knighter, DBLP:conf/emnlp/WangZSX024, DBLP:conf/nips/WangZSXX024, 10.1145/3649828}, typically following two directions.
One uses LLMs to supply domain knowledge like function specifications~\cite{DBLP:journals/pacmse/WangZW024, DBLP:journals/corr/abs-2405-17238, ye2024threads} and bug definitions~\cite{yang2025knighter}, to assist symbolic approaches in analyzing the program for bug detection. For example, IRIS leverages LLM-inferred sources and sinks to detect taint-style vulnerabilities~\cite{DBLP:journals/corr/abs-2405-17238}, while KNighter synthesizes static analyzers from bug patches for kernel bug detection~\cite{yang2025knighter}.
The other direction adopts agent-centric solutions that reason over source code directly without compilation~\cite{DBLP:conf/emnlp/WangZSX024, DBLP:conf/nips/WangZSXX024, 10.1145/3649828}.
Techniques such as LLMSAN~\cite{DBLP:conf/emnlp/WangZSX024} and LLMDFA~\cite{DBLP:conf/nips/WangZSXX024} combine LLMs with SMT solvers and parsing-based analyzers, while LLift uses progressive prompting to find kernel vulnerabilities~\cite{10.1145/3649828}. More recently, RepoAudit extends these efforts to detect multiple types of data-flow bugs~\cite{guo2025repoaudit}.
In contrast, our work targets functional bugs, where bug specifications often lack clear formalization. The diverse nature of functional properties also prevents the application of standard static analysis frameworks.
Nonetheless, our retrieval-based approach draws inspiration from traditional bug detection methodologies, enabling localizing code segments relevant to targeted functional properties, facilitating effective subsequent detection.
\section{Conclusion}\label{sec:conclusion}
This paper presents \sysname{}, an LLM agent designed for detecting functional bugs in network protocol implementations by identifying semantic inconsistencies between the code and RFC documents.
Specifically, \sysname{} comprises two complementary agents: an indexing agent responsible for semantic indexing of source code, and a detection agent performing retrieval-guided inconsistency detection.
In our evaluation, \sysname{} successfully uncovered 47 functional bugs across six different protocols, of which 20 have been acknowledged or fixed by the original developers. As the first LLM-based approach explicitly targeting functional bug detection in network protocols, our work offers valuable insights into future research on functional bug detection in domain-specific software systems, demonstrating the significant potential of leveraging LLM capabilities in security auditing tasks.

\section*{Acknowledgment}
We thank all the anonymous reviewers for the insightful and constructive feedback.
We are grateful to the Center for AI Safety for providing computational resources. This work was funded in part by the National Science Foundation (NSF) Awards SHF-1901242, SHF-1910300, Proto-OKN 2333736, IIS-2416835, DARPA VSPELLS - HR001120S0058, ONR N00014-23-1-2081, and Amazon. Any opinions, findings and conclusions or recommendations expressed in this material are those of the authors and do not necessarily reflect the views of the sponsors.

\bibliographystyle{IEEEtran}
\bibliography{ref}

\begin{thebibliography}{10}
\providecommand{\url}[1]{#1}
\csname url@samestyle\endcsname
\providecommand{\newblock}{\relax}
\providecommand{\bibinfo}[2]{#2}
\providecommand{\BIBentrySTDinterwordspacing}{\spaceskip=0pt\relax}
\providecommand{\BIBentryALTinterwordstretchfactor}{4}
\providecommand{\BIBentryALTinterwordspacing}{\spaceskip=\fontdimen2\font plus
\BIBentryALTinterwordstretchfactor\fontdimen3\font minus \fontdimen4\font\relax}
\providecommand{\BIBforeignlanguage}[2]{{%
\expandafter\ifx\csname l@#1\endcsname\relax
\typeout{** WARNING: IEEEtran.bst: No hyphenation pattern has been}%
\typeout{** loaded for the language `#1'. Using the pattern for}%
\typeout{** the default language instead.}%
\else
\language=\csname l@#1\endcsname
\fi
#2}}
\providecommand{\BIBdecl}{\relax}
\BIBdecl

\bibitem{Zerologon}
MITRE, ``Cve-2020-1472,'' \url{https://nvd.nist.gov/vuln/detail/CVE-2020-1472}, 2020.

\bibitem{MS-NRPC}
Microsoft, ``Netlogon remote protocol,'' \url{https://learn.microsoft.com/en-us/openspecs/windows_protocols/ms-nrpc/ff8f970f-3e37-40f7-bd4b-af7336e4792f}, 2024.

\bibitem{DBLP:journals/corr/AmorimHP17}
\BIBentryALTinterwordspacing
A.~A. de~Amorim, C.~Hritcu, and B.~C. Pierce, ``The meaning of memory safety,'' \emph{CoRR}, vol. abs/1705.07354, 2017. [Online]. Available: \url{http://arxiv.org/abs/1705.07354}
\BIBentrySTDinterwordspacing

\bibitem{LTLFuzz}
\BIBentryALTinterwordspacing
R.~Meng, Z.~Dong, J.~Li, I.~Beschastnikh, and A.~Roychoudhury, ``Linear-time temporal logic guided greybox fuzzing,'' in \emph{44th {IEEE/ACM} 44th International Conference on Software Engineering, {ICSE} 2022, Pittsburgh, PA, USA, May 25-27, 2022}.\hskip 1em plus 0.5em minus 0.4em\relax {ACM}, 2022, pp. 1343--1355. [Online]. Available: \url{https://doi.org/10.1145/3510003.3510082}
\BIBentrySTDinterwordspacing

\bibitem{feng2024rocas}
\BIBentryALTinterwordspacing
S.~Feng, Y.~Ye, Q.~Shi, Z.~Cheng, X.~Xu, S.~Cheng, H.~Choi, and X.~Zhang, ``{ROCAS:} root cause analysis of autonomous driving accidents via cyber-physical co-mutation,'' in \emph{Proceedings of the 39th {IEEE/ACM} International Conference on Automated Software Engineering, {ASE} 2024, Sacramento, CA, USA, October 27 - November 1, 2024}, V.~Filkov, B.~Ray, and M.~Zhou, Eds.\hskip 1em plus 0.5em minus 0.4em\relax {ACM}, 2024, pp. 1620--1632. [Online]. Available: \url{https://doi.org/10.1145/3691620.3695530}
\BIBentrySTDinterwordspacing

\bibitem{feng2025intentest}
\BIBentryALTinterwordspacing
S.~Feng, X.~Xu, X.~Chen, K.~Zhang, S.~Y. Ahmed, Z.~Su, M.~Zheng, and X.~Zhang, ``Intentest: Stress testing for intent integrity in api-calling {LLM} agents,'' \emph{CoRR}, vol. abs/2506.07524, 2025. [Online]. Available: \url{https://doi.org/10.48550/arXiv.2506.07524}
\BIBentrySTDinterwordspacing

\bibitem{kate2025roscallbax}
\BIBentryALTinterwordspacing
S.~Kate, Y.~Gao, S.~Feng, and X.~Zhang, ``Roscallbax: Statically detecting inconsistencies in callback function setup of robotic systems,'' \emph{Proc. {ACM} Softw. Eng.}, vol.~2, no. {FSE}, pp. 668--689, 2025. [Online]. Available: \url{https://doi.org/10.1145/3715748}
\BIBentrySTDinterwordspacing

\bibitem{klee}
\BIBentryALTinterwordspacing
C.~Cadar, D.~Dunbar, and D.~R. Engler, ``Klee: Unassisted and automatic generation of high-coverage tests for complex systems programs,'' in \emph{Proceedings of the 8th USENIX Symposium on Operating Systems Design and Implementation}, ser. OSDI '08.\hskip 1em plus 0.5em minus 0.4em\relax USENIX, 2008, pp. 209--224. [Online]. Available: \url{https://www.usenix.org/conference/osdi-08/klee-unassisted-and-automatic-generation-high-coverage-tests-complex-systems}
\BIBentrySTDinterwordspacing

\bibitem{255310}
\BIBentryALTinterwordspacing
S.~Poeplau and A.~Francillon, ``Symbolic execution with {SymCC}: Don{\textquoteright}t interpret, compile!'' in \emph{29th USENIX Security Symposium (USENIX Security 20)}.\hskip 1em plus 0.5em minus 0.4em\relax USENIX Association, Aug. 2020, pp. 181--198. [Online]. Available: \url{https://www.usenix.org/conference/usenixsecurity20/presentation/poeplau}
\BIBentrySTDinterwordspacing

\bibitem{217563}
\BIBentryALTinterwordspacing
I.~Yun, S.~Lee, M.~Xu, Y.~Jang, and T.~Kim, ``{QSYM} : A practical concolic execution engine tailored for hybrid fuzzing,'' in \emph{27th USENIX Security Symposium (USENIX Security 18)}.\hskip 1em plus 0.5em minus 0.4em\relax Baltimore, MD: USENIX Association, Aug. 2018, pp. 745--761. [Online]. Available: \url{https://www.usenix.org/conference/usenixsecurity18/presentation/yun}
\BIBentrySTDinterwordspacing

\bibitem{SFA-Miner}
J.~Jiang, M.~Zheng, Q.~Shi, and X.~Z. Zhang, ``{SFA-Miner}: Mining path-sensitive api usage patterns via symbolic finite automata,'' in \emph{The IEEE Symposium on Security and Privacy}, ser. S\&P 2026, 2026.

\bibitem{TensileFuzz}
\BIBentryALTinterwordspacing
X.~Liu, W.~You, Z.~Zhang, and X.~Zhang, ``Tensilefuzz: facilitating seed input generation in fuzzing via string constraint solving,'' in \emph{Proceedings of the 31st ACM SIGSOFT International Symposium on Software Testing and Analysis}, ser. ISSTA 2022.\hskip 1em plus 0.5em minus 0.4em\relax New York, NY, USA: Association for Computing Machinery, 2022, p. 391–403. [Online]. Available: \url{https://doi.org/10.1145/3533767.3534403}
\BIBentrySTDinterwordspacing

\bibitem{10.1145/3656400}
\BIBentryALTinterwordspacing
P.~Yao, J.~Zhou, X.~Xiao, Q.~Shi, R.~Wu, and C.~Zhang, ``Falcon: A fused approach to path-sensitive sparse data dependence analysis,'' \emph{Proc. ACM Program. Lang.}, vol.~8, no. PLDI, Jun. 2024. [Online]. Available: \url{https://doi.org/10.1145/3656400}
\BIBentrySTDinterwordspacing

\bibitem{10.1145/3180155.3180178}
\BIBentryALTinterwordspacing
H.~Yan, Y.~Sui, S.~Chen, and J.~Xue, ``Spatio-temporal context reduction: a pointer-analysis-based static approach for detecting use-after-free vulnerabilities,'' in \emph{Proceedings of the 40th International Conference on Software Engineering}, ser. ICSE '18.\hskip 1em plus 0.5em minus 0.4em\relax New York, NY, USA: Association for Computing Machinery, 2018, p. 327–337. [Online]. Available: \url{https://doi.org/10.1145/3180155.3180178}
\BIBentrySTDinterwordspacing

\bibitem{chatafl}
\BIBentryALTinterwordspacing
R.~Meng, M.~Mirchev, M.~B{\"{o}}hme, and A.~Roychoudhury, ``Large language model guided protocol fuzzing,'' in \emph{31st Annual Network and Distributed System Security Symposium, {NDSS} 2024, San Diego, California, USA, February 26 - March 1, 2024}.\hskip 1em plus 0.5em minus 0.4em\relax The Internet Society, 2024. [Online]. Available: \url{https://www.ndss-symposium.org/ndss-paper/large-language-model-guided-protocol-fuzzing/}
\BIBentrySTDinterwordspacing

\bibitem{mGPTFuzz}
\BIBentryALTinterwordspacing
X.~Ma, L.~Luo, and Q.~Zeng, ``From one thousand pages of specification to unveiling hidden bugs: Large language model assisted fuzzing of matter iot devices,'' in \emph{33rd {USENIX} Security Symposium, {USENIX} Security 2024, Philadelphia, PA, USA, August 14-16, 2024}, D.~Balzarotti and W.~Xu, Eds.\hskip 1em plus 0.5em minus 0.4em\relax {USENIX} Association, 2024. [Online]. Available: \url{https://www.usenix.org/conference/usenixsecurity24/presentation/ma-xiaoyue}
\BIBentrySTDinterwordspacing

\bibitem{FuzzInMem}
\BIBentryALTinterwordspacing
X.~Liu, W.~You, Y.~Ye, Z.~Zhang, J.~Huang, and X.~Zhang, ``Fuzzinmem: Fuzzing programs via in-memory structures,'' in \emph{Proceedings of the IEEE/ACM 46th International Conference on Software Engineering}, ser. ICSE '24.\hskip 1em plus 0.5em minus 0.4em\relax New York, NY, USA: Association for Computing Machinery, 2024. [Online]. Available: \url{https://doi.org/10.1145/3597503.3639172}
\BIBentrySTDinterwordspacing

\bibitem{Pardiff}
\BIBentryALTinterwordspacing
M.~Zheng, Q.~Shi, X.~Liu, X.~Xu, L.~Yu, C.~Liu, G.~Wei, and X.~Zhang, ``Pardiff: Practical static differential analysis of network protocol parsers,'' \emph{Proc. {ACM} Program. Lang.}, vol.~8, no. {OOPSLA1}, pp. 1208--1234, 2024. [Online]. Available: \url{https://doi.org/10.1145/3649854}
\BIBentrySTDinterwordspacing

\bibitem{ODIT}
\BIBentryALTinterwordspacing
R.~Rutledge and A.~Orso, ``Automating differential testing with overapproximate symbolic execution,'' in \emph{15th {IEEE} Conference on Software Testing, Verification and Validation, {ICST} 2022, Valencia, Spain, April 4-14, 2022}.\hskip 1em plus 0.5em minus 0.4em\relax {IEEE}, 2022, pp. 256--266. [Online]. Available: \url{https://doi.org/10.1109/ICST53961.2022.00035}
\BIBentrySTDinterwordspacing

\bibitem{reen2020dpifuzz}
\BIBentryALTinterwordspacing
G.~S. Reen and C.~Rossow, ``Dpifuzz: {A} differential fuzzing framework to detect {DPI} elusion strategies for {QUIC},'' in \emph{{ACSAC} '20: Annual Computer Security Applications Conference, Virtual Event / Austin, TX, USA, 7-11 December, 2020}.\hskip 1em plus 0.5em minus 0.4em\relax {ACM}, 2020, pp. 332--344. [Online]. Available: \url{https://doi.org/10.1145/3427228.3427662}
\BIBentrySTDinterwordspacing

\bibitem{Pistachio}
\BIBentryALTinterwordspacing
O.~Udrea and C.~Lumezanu, ``Rule-based static analysis of network protocol implementations,'' in \emph{Proceedings of the 15th {USENIX} Security Symposium, Vancouver, BC, Canada, July 31 - August 4, 2006}, A.~D. Keromytis, Ed.\hskip 1em plus 0.5em minus 0.4em\relax {USENIX} Association, 2006. [Online]. Available: \url{https://www.usenix.org/conference/15th-usenix-security-symposium/rule-based-static-analysis-network-protocol}
\BIBentrySTDinterwordspacing

\bibitem{MusuvathiE04}
\BIBentryALTinterwordspacing
M.~Musuvathi and D.~R. Engler, ``Model checking large network protocol implementations,'' in \emph{1st Symposium on Networked Systems Design and Implementation {(NSDI} 2004), March 29-31, 2004, San Francisco, California, USA, Proceedings}, R.~Morris and S.~Savage, Eds.\hskip 1em plus 0.5em minus 0.4em\relax {USENIX}, 2004, pp. 155--168. [Online]. Available: \url{http://www.usenix.org/events/nsdi04/tech/musuvathi.html}
\BIBentrySTDinterwordspacing

\bibitem{rfctool}
``The artifact of {RFCAudit},'' \url{https://github.com/zmw12306/RFCAudit}, 2025.

\bibitem{KIT}
\BIBentryALTinterwordspacing
C.~Liu, S.~Gong, and P.~Fonseca, ``{KIT:} testing os-level virtualization for functional interference bugs,'' in \emph{Proceedings of the 28th {ACM} International Conference on Architectural Support for Programming Languages and Operating Systems, Volume 2, {ASPLOS} 2023, Vancouver, BC, Canada, March 25-29, 2023}, T.~M. Aamodt, N.~D.~E. Jerger, and M.~M. Swift, Eds.\hskip 1em plus 0.5em minus 0.4em\relax {ACM}, 2023, pp. 427--441. [Online]. Available: \url{https://doi.org/10.1145/3575693.3575731}
\BIBentrySTDinterwordspacing

\bibitem{pham2020aflnet}
\BIBentryALTinterwordspacing
V.~Pham, M.~B{\"{o}}hme, and A.~Roychoudhury, ``{AFLNET:} {A} greybox fuzzer for network protocols,'' in \emph{13th {IEEE} International Conference on Software Testing, Validation and Verification, {ICST} 2020, Porto, Portugal, October 24-28, 2020}.\hskip 1em plus 0.5em minus 0.4em\relax {IEEE}, 2020, pp. 460--465. [Online]. Available: \url{https://doi.org/10.1109/ICST46399.2020.00062}
\BIBentrySTDinterwordspacing

\bibitem{shi2023lifting}
\BIBentryALTinterwordspacing
Q.~Shi, J.~Shao, Y.~Ye, M.~Zheng, and X.~Zhang, ``Lifting network protocol implementation to precise format specification with security applications,'' in \emph{Proceedings of the 2023 {ACM} {SIGSAC} Conference on Computer and Communications Security, {CCS} 2023, Copenhagen, Denmark, November 26-30, 2023}, W.~Meng, C.~D. Jensen, C.~Cremers, and E.~Kirda, Eds.\hskip 1em plus 0.5em minus 0.4em\relax {ACM}, 2023, pp. 1287--1301. [Online]. Available: \url{https://doi.org/10.1145/3576915.3616614}
\BIBentrySTDinterwordspacing

\bibitem{shi2018pinpoint}
\BIBentryALTinterwordspacing
Q.~Shi, X.~Xiao, R.~Wu, J.~Zhou, G.~Fan, and C.~Zhang, ``Pinpoint: fast and precise sparse value flow analysis for million lines of code,'' in \emph{Proceedings of the 39th {ACM} {SIGPLAN} Conference on Programming Language Design and Implementation, {PLDI} 2018, Philadelphia, PA, USA, June 18-22, 2018}, J.~S. Foster and D.~Grossman, Eds.\hskip 1em plus 0.5em minus 0.4em\relax {ACM}, 2018, pp. 693--706. [Online]. Available: \url{https://doi.org/10.1145/3192366.3192418}
\BIBentrySTDinterwordspacing

\bibitem{Xue16SVF}
\BIBentryALTinterwordspacing
Y.~Sui and J.~Xue, ``{SVF:} interprocedural static value-flow analysis in {LLVM},'' in \emph{Proceedings of the 25th International Conference on Compiler Construction, {CC} 2016, Barcelona, Spain, March 12-18, 2016}, A.~Zaks and M.~V. Hermenegildo, Eds.\hskip 1em plus 0.5em minus 0.4em\relax {ACM}, 2016, pp. 265--266. [Online]. Available: \url{https://doi.org/10.1145/2892208.2892235}
\BIBentrySTDinterwordspacing

\bibitem{DiazCRP04}
\BIBentryALTinterwordspacing
G.~D{\'{\i}}az, F.~Cuartero, V.~V. Ruiz, and F.~L. Pelayo, ``Automatic verification of the {TLS} handshake protocol,'' in \emph{Proceedings of the 2004 {ACM} Symposium on Applied Computing (SAC), Nicosia, Cyprus, March 14-17, 2004}, H.~Haddad, A.~Omicini, R.~L. Wainwright, and L.~M. Liebrock, Eds.\hskip 1em plus 0.5em minus 0.4em\relax {ACM}, 2004, pp. 789--794. [Online]. Available: \url{https://doi.org/10.1145/967900.968063}
\BIBentrySTDinterwordspacing

\bibitem{ParCleanse}
\BIBentryALTinterwordspacing
M.~Zheng, D.~Xie, Q.~Shi, C.~Wang, and X.~Zhang, ``Validating network protocol parsers with traceable {RFC} document interpretation,'' \emph{Proc. {ACM} Softw. Eng.}, vol.~2, no. {ISSTA}, pp. 1772--1794, 2025. [Online]. Available: \url{https://doi.org/10.1145/3728955}
\BIBentrySTDinterwordspacing

\bibitem{chen2025locagent}
\BIBentryALTinterwordspacing
Z.~Chen, R.~Tang, G.~Deng, F.~Wu, J.~Wu, Z.~Jiang, V.~K. Prasanna, A.~Cohan, and X.~Wang, ``Locagent: Graph-guided {LLM} agents for code localization,'' in \emph{Proceedings of the 63rd Annual Meeting of the Association for Computational Linguistics (Volume 1: Long Papers), {ACL} 2025, Vienna, Austria, July 27 - August 1, 2025}, W.~Che, J.~Nabende, E.~Shutova, and M.~T. Pilehvar, Eds.\hskip 1em plus 0.5em minus 0.4em\relax Association for Computational Linguistics, 2025, pp. 8697--8727. [Online]. Available: \url{https://aclanthology.org/2025.acl-long.426/}
\BIBentrySTDinterwordspacing

\bibitem{xia2024agentless}
\BIBentryALTinterwordspacing
C.~S. Xia, Y.~Deng, S.~Dunn, and L.~Zhang, ``Demystifying llm-based software engineering agents,'' \emph{Proc. {ACM} Softw. Eng.}, vol.~2, no. {FSE}, pp. 801--824, 2025. [Online]. Available: \url{https://doi.org/10.1145/3715754}
\BIBentrySTDinterwordspacing

\bibitem{lin-etal-2024-criticbench}
\BIBentryALTinterwordspacing
Z.~Lin, Z.~Gou, T.~Liang, R.~Luo, H.~Liu, and Y.~Yang, ``{C}ritic{B}ench: Benchmarking {LLM}s for critique-correct reasoning,'' in \emph{Findings of the Association for Computational Linguistics: ACL 2024}, L.-W. Ku, A.~Martins, and V.~Srikumar, Eds.\hskip 1em plus 0.5em minus 0.4em\relax Bangkok, Thailand: Association for Computational Linguistics, Aug. 2024, pp. 1552--1587. [Online]. Available: \url{https://aclanthology.org/2024.findings-acl.91/}
\BIBentrySTDinterwordspacing

\bibitem{AutoGen}
\BIBentryALTinterwordspacing
Q.~Wu, G.~Bansal, J.~Zhang, Y.~Wu, B.~Li, E.~Zhu, L.~Jiang, X.~Zhang, S.~Zhang, J.~Liu, A.~H. Awadallah, R.~W. White, D.~Burger, and C.~Wang, ``Autogen: Enabling next-gen {LLM} applications via multi-agent conversation,'' in \emph{ICLR 2024 Workshop on Large Language Model (LLM) Agents}, 2024. [Online]. Available: \url{https://openreview.net/forum?id=uAjxFFing2}
\BIBentrySTDinterwordspacing

\bibitem{claude}
Anthropic, ``Claude 3.5 sonnet,'' \url{https://www.anthropic.com/claude/sonnet}, 2025.

\bibitem{treesitter}
``Tree-sitter,'' \url{https://tree-sitter.github.io/tree-sitter/}.

\bibitem{frr}
F.~community, ``The frrouting protocol suite,'' \url{https://github.com/FRRouting/frr}, 2024.

\bibitem{lwip}
``lwip - a lightweight tcp/ip stack,'' \url{https://github.com/lwip-tcpip/lwip}, 2025.

\bibitem{ParVal}
\BIBentryALTinterwordspacing
M.~Zheng, D.~Xie, and X.~Zhang, ``Large language models for validating network protocol parsers,'' in \emph{2025 {IEEE} Security and Privacy, {SP} 2025 - Workshops, San Francisco, CA, USA, May 15, 2025}, M.~Blanton, W.~Enck, and C.~Nita{-}Rotaru, Eds.\hskip 1em plus 0.5em minus 0.4em\relax {IEEE}, 2025, pp. 56--64. [Online]. Available: \url{https://doi.org/10.1109/SPW67851.2025.00009}
\BIBentrySTDinterwordspacing

\bibitem{deng2025swe}
\BIBentryALTinterwordspacing
X.~Deng, J.~Da, E.~Pan, Y.~Y. He, C.~Ide, K.~Garg, N.~Lauffer, A.~Park, N.~Pasari, C.~Rane \emph{et~al.}, ``Swe-bench pro: Can ai agents solve long-horizon software engineering tasks?'' \emph{arXiv preprint arXiv:2509.16941}, 2025. [Online]. Available: \url{https://arxiv.org/abs/2509.16941}
\BIBentrySTDinterwordspacing

\bibitem{xie2023impact}
\BIBentryALTinterwordspacing
D.~Xie, B.~Yoo, N.~Jiang, M.~Kim, L.~Tan, X.~Zhang, and J.~S. Lee, ``Impact of large language models on generating software specifications,'' \emph{CoRR}, vol. abs/2306.03324, 2023. [Online]. Available: \url{https://doi.org/10.48550/arXiv.2306.03324}
\BIBentrySTDinterwordspacing

\bibitem{xie2025core}
\BIBentryALTinterwordspacing
D.~Xie, M.~Zheng, X.~Liu, J.~Wang, C.~Wang, L.~Tan, and X.~Zhang, ``Core: Benchmarking {LLM}s{\textquoteright} code reasoning capabilities through static analysis tasks,'' in \emph{The Thirty-ninth Annual Conference on Neural Information Processing Systems Datasets and Benchmarks Track}, 2025. [Online]. Available: \url{https://openreview.net/forum?id=WJIDorHiuZ}
\BIBentrySTDinterwordspacing

\bibitem{9678600}
\BIBentryALTinterwordspacing
M.~Zheng, J.~Yang, M.~Wen, H.~Zhu, Y.~Liu, and H.~Jin, ``Why do developers remove lambda expressions in java?'' in \emph{36th {IEEE/ACM} International Conference on Automated Software Engineering, {ASE} 2021, Melbourne, Australia, November 15-19, 2021}.\hskip 1em plus 0.5em minus 0.4em\relax {IEEE}, 2021, pp. 67--78. [Online]. Available: \url{https://doi.org/10.1109/ASE51524.2021.9678600}
\BIBentrySTDinterwordspacing

\bibitem{boofuzz}
J.~Pereyda, ``Boofuzz,'' \url{https://github.com/jtpereyda/boofuzz}, 2023.

\bibitem{Msrsage}
\BIBentryALTinterwordspacing
P.~Godefroid, M.~Y. Levin, and D.~A. Molnar, ``{SAGE:} whitebox fuzzing for security testing,'' \emph{Commun. {ACM}}, vol.~55, no.~3, pp. 40--44, 2012. [Online]. Available: \url{https://doi.org/10.1145/2093548.2093564}
\BIBentrySTDinterwordspacing

\bibitem{wang2024llmif}
\BIBentryALTinterwordspacing
J.~Wang, L.~Yu, and X.~Luo, ``{LLMIF:} augmented large language model for fuzzing iot devices,'' in \emph{{IEEE} Symposium on Security and Privacy, {SP} 2024, San Francisco, CA, USA, May 19-23, 2024}.\hskip 1em plus 0.5em minus 0.4em\relax {IEEE}, 2024, pp. 881--896. [Online]. Available: \url{https://doi.org/10.1109/SP54263.2024.00211}
\BIBentrySTDinterwordspacing

\bibitem{EBugDec}
\BIBentryALTinterwordspacing
J.~Chen, F.~Li, Q.~Chen, P.~Li, L.~Xu, and W.~Huo, ``Ebugdec: Detecting inconsistency bugs caused by {RFC} evolution in protocol implementations,'' in \emph{Proceedings of the 26th International Symposium on Research in Attacks, Intrusions and Defenses, {RAID} 2023, Hong Kong, China, October 16-18, 2023}.\hskip 1em plus 0.5em minus 0.4em\relax {ACM}, 2023, pp. 412--425. [Online]. Available: \url{https://doi.org/10.1145/3607199.3607222}
\BIBentrySTDinterwordspacing

\bibitem{DBLP:journals/pacmse/WangZW024}
\BIBentryALTinterwordspacing
C.~Wang, J.~Zhang, R.~Wu, and C.~Zhang, ``Dainfer: Inferring {API} aliasing specifications from library documentation via neurosymbolic optimization,'' \emph{Proc. {ACM} Softw. Eng.}, vol.~1, no. {FSE}, pp. 2469--2492, 2024. [Online]. Available: \url{https://doi.org/10.1145/3660816}
\BIBentrySTDinterwordspacing

\bibitem{ye2024threads}
\BIBentryALTinterwordspacing
C.~Ye, Y.~Cai, and C.~Zhang, ``When threads meet interrupts: Effective static detection of interrupt-based deadlocks in linux,'' in \emph{33rd {USENIX} Security Symposium, {USENIX} Security 2024, Philadelphia, PA, USA, August 14-16, 2024}, D.~Balzarotti and W.~Xu, Eds.\hskip 1em plus 0.5em minus 0.4em\relax {USENIX} Association, 2024. [Online]. Available: \url{https://www.usenix.org/conference/usenixsecurity24/presentation/ye}
\BIBentrySTDinterwordspacing

\bibitem{DBLP:journals/corr/abs-2405-17238}
\BIBentryALTinterwordspacing
Z.~Li, S.~Dutta, and M.~Naik, ``Llm-assisted static analysis for detecting security vulnerabilities,'' \emph{CoRR}, vol. abs/2405.17238, 2024. [Online]. Available: \url{https://doi.org/10.48550/arXiv.2405.17238}
\BIBentrySTDinterwordspacing

\bibitem{yang2025knighter}
\BIBentryALTinterwordspacing
C.~Yang, Z.~Zhao, Z.~Xie, H.~Li, and L.~Zhang, ``Knighter: Transforming static analysis with llm-synthesized checkers,'' \emph{CoRR}, vol. abs/2503.09002, 2025. [Online]. Available: \url{https://doi.org/10.48550/arXiv.2503.09002}
\BIBentrySTDinterwordspacing

\bibitem{DBLP:conf/emnlp/WangZSX024}
\BIBentryALTinterwordspacing
C.~Wang, W.~Zhang, Z.~Su, X.~Xu, and X.~Zhang, ``Sanitizing large language models in bug detection with data-flow,'' in \emph{Findings of the Association for Computational Linguistics: {EMNLP} 2024, Miami, Florida, USA, November 12-16, 2024}, Y.~Al{-}Onaizan, M.~Bansal, and Y.~Chen, Eds.\hskip 1em plus 0.5em minus 0.4em\relax Association for Computational Linguistics, 2024, pp. 3790--3805. [Online]. Available: \url{https://aclanthology.org/2024.findings-emnlp.217}
\BIBentrySTDinterwordspacing

\bibitem{DBLP:conf/nips/WangZSXX024}
\BIBentryALTinterwordspacing
C.~Wang, W.~Zhang, Z.~Su, X.~Xu, X.~Xie, and X.~Zhang, ``{LLMDFA:} analyzing dataflow in code with large language models,'' in \emph{Advances in Neural Information Processing Systems 38: Annual Conference on Neural Information Processing Systems 2024, NeurIPS 2024, Vancouver, BC, Canada, December 10 - 15, 2024}, A.~Globersons, L.~Mackey, D.~Belgrave, A.~Fan, U.~Paquet, J.~M. Tomczak, and C.~Zhang, Eds., 2024. [Online]. Available: \url{http://papers.nips.cc/paper\_files/paper/2024/hash/ed9dcde1eb9c597f68c1d375bbecf3fc-Abstract-Conference.html}
\BIBentrySTDinterwordspacing

\bibitem{10.1145/3649828}
\BIBentryALTinterwordspacing
H.~Li, Y.~Hao, Y.~Zhai, and Z.~Qian, ``Enhancing static analysis for practical bug detection: An llm-integrated approach,'' \emph{Proc. ACM Program. Lang.}, vol.~8, no. OOPSLA1, Apr. 2024. [Online]. Available: \url{https://doi.org/10.1145/3649828}
\BIBentrySTDinterwordspacing

\bibitem{guo2025repoaudit}
\BIBentryALTinterwordspacing
J.~Guo, C.~Wang, X.~Xu, Z.~Su, and X.~Zhang, ``Repoaudit: An autonomous llm-agent for repository-level code auditing,'' \emph{CoRR}, vol. abs/2501.18160, 2025. [Online]. Available: \url{https://doi.org/10.48550/arXiv.2501.18160}
\BIBentrySTDinterwordspacing

\end{thebibliography}
\end{document}